\documentclass[twocolumn,usegraphicx]{emulateapj}
\usepackage{natbib}
\usepackage{epstopdf}
\usepackage{amsmath}
\usepackage{physics}
\usepackage{float}
\usepackage{enumitem}
\usepackage[backref,breaklinks,colorlinks,citecolor=cyan]{hyperref}
\usepackage[all]{hypcap} 

\shorttitle{Disentangling Doppler signals and stellar activity}
\shortauthors{D\'iaz et al.}

\begin{document}

\title{The test case of HD26965: difficulties disentangling weak Doppler signals from stellar activity$^{7}$}

\author{Mat\'ias R. D\'iaz$^{1,2,\star}$, 
James S. Jenkins$^{1,3}$,
Mikko Tuomi$^{4}$,
R. Paul Butler$^{5}$,
Maritza G. Soto$^{1}$,\\
Johanna K. Teske$^{2,5}$,
Fabo Feng$^{4}$,
Stephen A. Shectman$^{2}$,
Pamela Arriagada$^{5}$,
Jeffrey D. Crane$^{2}$,
Ian B. Thompson$^{2}$
\& Steven S. Vogt$^{6}$
}
\altaffiltext{}
{\\
$^{1}$ Departamento de Astronom\'ia, Universidad de Chile, Camino El Observatorio 1515, Las Condes, Santiago, Chile.\\
$^{2}$ The Observatories, Carnegie Institution for Science, 813 Santa Barbara Street, Pasadena, CA 91101, USA.\\
$^{3}$ Centro de Astrof\'isica y Tecnolog\'ias Afines (CATA), Casilla 36-D, Santiago, Chile.\\
$^{4}$ Center for Astrophysics Research, School of Physics, Astronomy and Mathematics, University of Hertfordshire, College Lane, Hatfield AL109AB, UK.\\
$^{5}$ Department of Terrestrial Magnetism, Carnegie Institution of Washington, 5124 Broad Branch Road, Washington, DC 20015-1305, USA.\\
$^{6}$ UCO/Lick Observatory, Department of Astronomy and Astrophysics, University of California at Santa Cruz, Santa Cruz, CA 95064, USA.\\
$^{7}${ Some of the data presented herein were obtained at the W.M. Keck Observatory, which is operated as a scientific partnership among the California Institute of Technology, the University of California and the National Aeronautics and Space Administration. The Observatory was made possible by the generous financial support of the W.M. Keck Foundation.}\\
$^{\star}$\url{matias.diaz.m@ug.uchile.cl}, Carnegie-Chile Graduate Fellow 2015-2016.\\
}


\begin{abstract}
{ We report the discovery of a radial velocity signal that can be interpreted as a planetary-mass candidate orbiting the K dwarf HD26965, with an orbital period of 42.364$\pm$0.015 days, or alternatively, as the presence of residual, uncorrected rotational activity in the data. Observations include data from HIRES, PFS, CHIRON, and HARPS, where 1,111 measurements were made over 16 years. Our best solution for HD26965 $b$ is consistent with a super-Earth that has a minimum mass of 6.92$\pm$0.79 M$_{\oplus}$ orbiting at a distance of 0.215$\pm$0.008 AU from its host star.  We have analyzed the correlation between spectral activity indicators and the radial velocities from each instrument, showing moderate correlations that we include in our model. From this analysis, we recover a $\sim$38 day signal, which matches some literature values of the stellar rotation period. However, from independent Mt. Wilson HK data for this star, we find evidence for a significant 42 day signal after subtraction of longer period magnetic cycles, casting doubt on the planetary hypothesis for this period. Although our statistical model strongly suggests that the 42-day signal is Doppler in origin, we conclude that the residual effects of stellar rotation are difficult to fully model and remove from this dataset, highlighting the difficulties to disentangle small planetary signals and photospheric noise, particularly when the orbital periods are close to the rotation period of the star. This study serves as an excellent test case for future works that aim to detect small planets orbiting `Sun-like' stars using radial velocity measurements.}
\end{abstract}

\keywords{stars: individual HD26965 --- techniques: spectroscopic, radial velocities --- methods: statistical}

\section{Introduction}
Planets orbiting the nearest stars to the Sun are the most highly prized of all exoplanets, since they represent the most accessible targets for follow-up characterization studies. The measurement of precision radial velocities has allowed us to begin to build up a collection of planets orbiting the nearest stars, while also characterizing their orbital parameters. In particular, discoveries like 51 Peg $b$ \citep{MayorQueloz1995}, 47 UMa $b$ \citep{ButlerMarcy1996}, 70 Vir $b$ \citep{MarcyButler1996}, HD143361 $b$ and HD154672 $b$ \citep{Jenkins2009}, HD86226 $b$, HD164604 $b$, HD175167 $b$ \citep{Arriagada2010}, HD128356 $b$, HD154672 $b$ and HD224538 $b$ \citep{Jenkins2017}, GJ 876 $b$, $c$, $d$, $e$ \citep{Rivera2010}, and $\upsilon$ And $b$, $c$, $d$ \citep{Wright2009,Curiel2011}, among others, have allowed us to explore the wide diversity of gas giant planetary systems.

In the last few years, the advances in radial velocity precision that have been driven by technology improvements and better analysis methods have allowed the discovery of the first batch of low-mass planets orbiting nearby stars, e.g., GJ 876 $d$ \citep{Rivera2010}, HD40307 $b$, $c$, $d$, $e$, $f$ and $g$ \citep{Mayor2009, Tuomi2013a}, GJ 581 $d$ \citep{Vogt2010}, GJ 667C $b$, $c$ and $d$ \citep{AngladaEscude2012, AngladaEscude2013}, the candidates orbiting $\tau$ Ceti, planets $b$, $c$, $d$, $e$ and $f$ \citep{Tuomi2013b, Feng2017} and more recently Proxima Centauri $b$ \citep{AngladaEscude2016} represent a new population of super-Earth planets not witnessed in the Solar System, and are defined as being small planets with masses $\sim$2-10$M_{\oplus}$ that can either be primarily rocky objects or more fluffy, atmosphere dominated worlds \citep{Valencia2007, Kaltenegger2011}. 

In comparison to the gas giants, super-Earths seem to have some dramatically different characteristics, likely related to their formation and early evolution. They generally appear to be orbiting on mostly circular orbits \citep{TuomiAngladaEscude2013}, come in tightly packed planetary systems \citep{Lissauer2011, Latham2011}, and do not seem to follow the same metallicity bias as the gas giants \citep{Buchhave2012, Courcol2016}. In fact, there may be a lack of low-mass planets orbiting nearby and super metal-rich Sun-like stars \citep{Jenkins2009, Jenkins2013}. Models that invoke core accretion as the dominant planet formation scenario predict some of these trends, with mass functions rising heavily toward the lowest masses \citep{Mordasini2008}, also shown by analysis of the radial velocity sample of detected planets \citep{LopezJenkins2012}. 
Planetary formation models also predict a damping of the metallicity bias in planet fraction for low-mass objects, since the stellar metallicity is an observational proxy of the dust content in the inner disk when the planets were undergoing formation. However the picture may be less clear, since \citet{Mulders2016} have shown that there might be an increase in the occurrence of small rocky planets around host stars with super-solar metallicities and orbital periods $<$ 10 days.

Although the radial velocity method has been very successful at planet detection, it is an indirect method and therefore care must be taken when trying to confirm any signal with an amplitude at the few m s$^{-1}$ level (like many super-Earth signals), since this is the domain where stellar activity effects that are correlated with the rotation of the star can impact the data \citep{Boisse2011, Boisse2012}. In numerous cases, both large and small planet candidates have been challenged as being due to the effects of stellar activity (e.g., HD166435 \citealt{Queloz2001}; HIP13044, \citealt{JonesJenkins2014}; HD41248, \citealt{Santos2014}; GJ 581 $d$, \citealt{Robertson2014}; Kapteyn $b$  \citealt{Robertson2015}; $\alpha$ Cen B $b$, \citealt{Rajpaul2016}), with most of these challenges leading to counter-claims (e.g. HD41248, \citealt{JenkinsTuomi2014}; GJ 581 $d$, \citealt{AngladaEscudeTuomi2015}; Kapteyn $b$ and $c$, \citealt{Anglada2016}). Therefore due care must be taken to ensure any signal has been well inspected for the effects of stellar activity and/or stellar rotation.

Once a planet has been confirmed orbiting a nearby star, there exists the ability to perform detailed secondary follow-up studies, like measuring accurate stellar atomic and molecular abundances \citep{Schuler2015, Melendez2017} that could be sign-posts of planetary systems, or searching for transits and secondary eclipse measurements \citep{Baskin2013,Chen2014,vonParis2016}.  The combination of mass from the velocities and radius from any detected transit allows the bulk density of the planet to be measured (e.g., BD+20594 $b$, \citealt{Espinoza2016}; GJ 1214 $b$, \citealt{Valencia2013}; GJ 436 $b$ \citealt{vonBraun2012, Lanotte2014}; 55 Cnc $e$, \citealt{deMooij2014, Winn2011}) and from there, model comparisons can be made to infer the bulk composition. Therefore, gaining a better understanding of the population of low-mass planets requires the detection of more of these worlds orbiting bright stars in the solar neighborhood.

Here we present data from a 16 year precision radial velocity monitoring campaign, using multiple high resolution optical spectrographs, of the nearby ($\sim$5 pc) K0.5 star HD26965.

\section{HD 26965 - Stellar properties}\label{sec:prop}
HD26965 (HIP19849, GJ 166A) is classified as a K0.5V star \citep{Gray2006} with a visual magnitude of $V$=4.43 and an optical color of $B-V$=$0.82$. An activity index of log$\,R_{\rm HK}^{'}$=$-4.99$ is reported by \citet{Jenkins2011}. This value is also consistent with measurements found in other sources in the literature (e.g., -5.09, \citealt{Gray2006}; -4.97, \citealt{Murgas2013}) and a comparison with the Sun's mean activity value of log$\,R_{\rm HK \odot}^{'}$=$-4.91$ \citep{MamajekHillenbrand2008} tells us that HD26965 is a chromospherically quiet star.

\begin{figure}\label{fig:specieschains}
\centering
\includegraphics[scale=0.47]{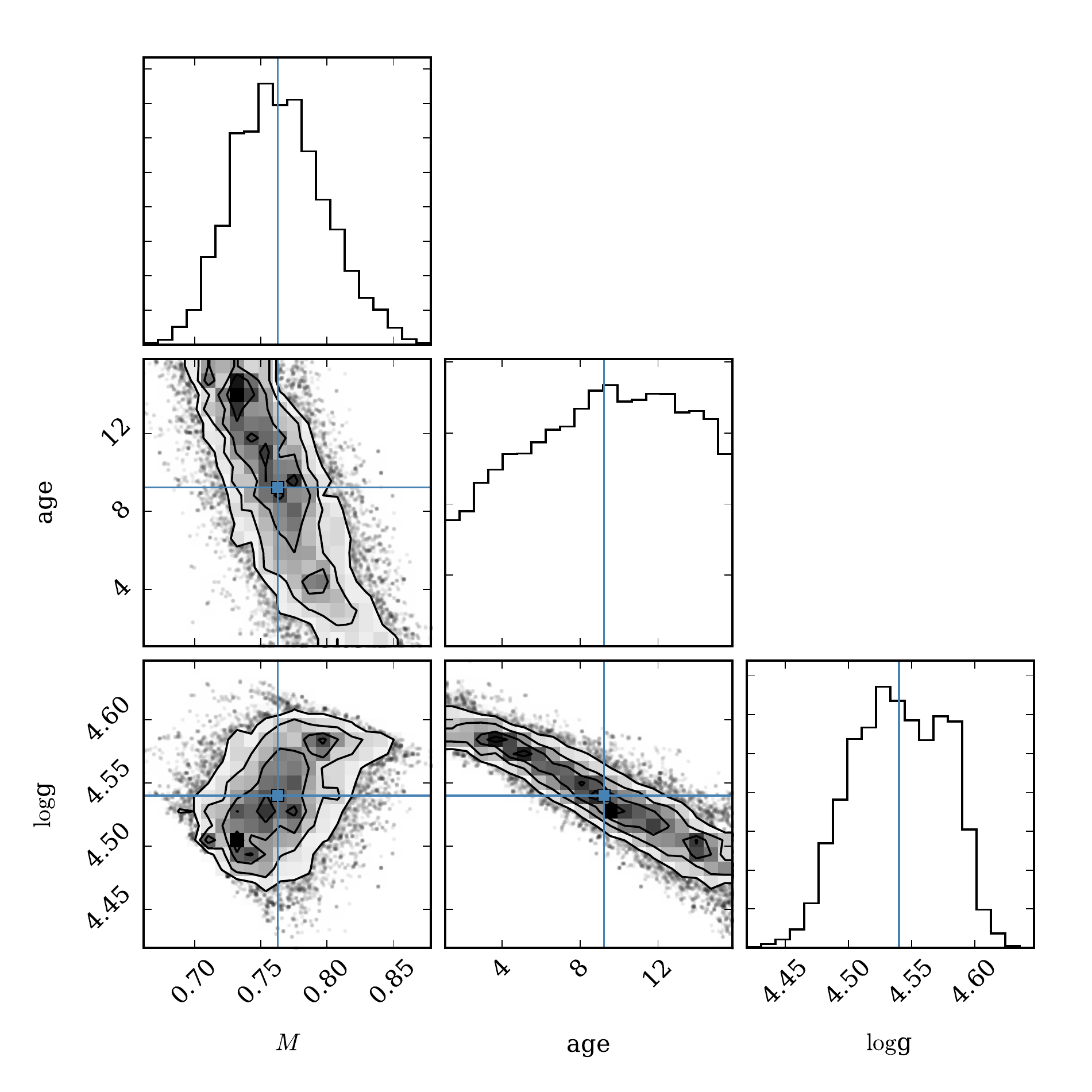}
\caption{Corner plot showing the one and two dimensional projections of the posterior probability distributions for the mass, age and log $g$ parameters estimated via MCMC samples with SPECIES. The plot has been generated using the Python pa\-ckage \texttt{corner.py} \citep{corner}.}
\end{figure}

The remaining stellar parameters were estimated using the {\bf S}pectroscopic {\bf P}arameters and atmosphe{\bf E}ric {\bf C}hem{\bf I}stri{\bf E}s of {\bf S}tars code ({SPECIES}; Soto \& Jenkins, submitted). {SPECIES} derives the effective temperature, surface gravity, metallicity, microturbulence, macroturbulence and rotational velocity, mass, age and chemical abundances for 11 elements in a self-consistent and automatic manner. The parameters were derived using high-resolution, high signal-to-noise spectra as input for the code, where in this particular case, we have used spectra from HARPS to derive the stellar parameters with SPECIES.  The first four parameters were found by measuring the equivalent widths (EWs) for a set of iron lines using the {ARES} code \citep{Sousa2007}. These values, along with a stellar atmosphere model \citep{Kurucz1993}, were then input to {MOOG} \citep{Sneden1973}, which solves the radiative transfer equation by imposing excitation and ionization equilibrium. Following on from this, we then derived the chemical abundances, measuring the EWs of a set of lines for each element. Macroturbulence and rotation velocity were computed by measuring the broadening of spectral lines using a Fourier analysis. Finally, mass and age were found by fitting isochrones \citep{Dotter2008} using the luminosity and temperature of the star. Figure \ref{fig:specieschains} shows the final distributions for the stellar mass, age and log $g$ using a Markov chain Monte Carlo (MCMC) method within SPECIES. More details about this code, in particular on the treatment of correlations between parameters and uncertainties, can be found in Soto \& Jenkins (submitted).

We found HD26965 to have a metallicity [Fe/H] of -0.29 $\pm$ 0.13 dex,  consistent with previously reported values (e.g., -0.28 dex \citealt{Gray2006, ValentiFischer2005, Turnbull2015}), and significantly poorer in metals than the Sun. SPECIES finds a rotational velocity of $v\,$sin$\,i$=1.23 $\pm$ 0.28 km s$^{-1}$, which is in agreement with the values of 1.1 $\pm$ 1.0, 1.4 $\pm$ 0.8, 1.6 $\pm$ 0.8 km s$^{-1}$ reported by \citet{Glebocki2005} calculated via cross-correlation, calibrated line full width at half maximum (FWHM), and convolution with rotational broadening profiles, respectively. 

The $v\,$sin$\,i$ value we report is consistent with the old age of the star found by {SPECIES} if we consider that stars on the main sequence spin-down with time due to the loss of angular momentum from winds and the increase in stellar radius with time that is required to maintain hydrostatic equilibrium as the core changes due to nuclear burning. The Sun has a rotational velocity of only 1.6 $\pm$ 0.3 km s$^{-1}$ \citep{Pavlenko2012} and we classify it as a slow rotator. In summary, the values found for the parameters make HD26965 a good candidate for radial velocity planet search since it can be considered a quiescent and slowly rotating star. The properties and derived parameters for HD26965 are summarized in Table \ref{tab:properties}.

\begin{table}\label{tab:properties}
\center
\caption{Stellar Parameters of HD26965.}
\label{tab:params}
\begin{tabular}{lcc}
\hline\hline
\multicolumn{1}{l}{Parameter}& \multicolumn{1}{c}{ Value  } & \multicolumn{1}{c}{Source}\\ \hline
R.A. (J2000)& 04:15:16.32  & SIMBAD\\
Dec. (J2000)& -07:39:10.34 & SIMBAD\\
$m_{V}$ & 4.43 & SIMBAD\\
$B$-$V$ & 0.82& SIMBAD \\
Distance (pc) & 4.98 $\pm$ 0.01 &    \citealt{vanLeeuwen2007}\\ \hline
Spectral type & K0.5V & \citealt{Gray2006} \\
Mass ($M_{\odot}$) & 0.76 $\pm$ 0.03 & This work (SPECIES)\\
Age (Gyr) & 9.23 $\pm$ 4.84 & This work (SPECIES)\\
Luminosity ($L_{\odot}$)& 0.44 & \citealt{Anderson2012}\\
T$_{\rm eff}$ (K) & 5151 $\pm$ 55 & This work (SPECIES)\\
\lbrack Fe/H\rbrack  &  -0.29 $\pm$ 0.12 & This work (SPECIES)\\
log $g$& $4.54 \pm$ 0.04 & This work (SPECIES)\\
$v$ sin $i$ (km s$^{-1}$) & 1.23 $\pm$ 0.28 & This work (SPECIES)\\
log $R^{'}_{HK}$ & -4.99 & \citealt{Jenkins2011}\\ \hline
\end{tabular}
\vspace{0.5cm}
\end{table}

\section{Spectroscopic Observations}\label{sec:obs}

High-precision Doppler measurements of HD26965 were carried out using 4 different spectrographs: The High Resolution Echelle Spectrograph (HIRES) installed on the 10 m Keck Telescope in Hawaii, the Carnegie Planet Finder Spectrograph (PFS) mounted on the 6.5 m Magellan II (Clay) telescope at Las Campanas Observatory, CHIRON mounted on the 1.5 m telescope from the Small to Moderate Aperture Research Telescopes (SMARTS) consortium in Cerro Tololo Interamerican Observatory and the High Accuracy Radial velocity Planet Searcher (HARPS) installed on the 3.6 m ESO telescope at La Silla Observatory.

\begin{table*}\label{tab:obs_params}
\center
\caption{Summary of instrumental and observational parameters for the different instruments.}
\label{tab:params}
\begin{tabular}{lcccccc}
\hline\hline
\multicolumn{1}{l}{Instrument}& \multicolumn{1}{c}{ Resolution  } & \multicolumn{1}{c}{$<$SNR$>$/Resol. element} &\multicolumn{1}{c}{$<$Exposure time$>$}&\multicolumn{1}{c}{$<$RV error$ >$}&\multicolumn{1}{c}{N$_{\rm obs}$}&\multicolumn{1}{c}{ Time baseline}\\
\multicolumn{1}{l}{}& \multicolumn{1}{c}{} & \multicolumn{1}{c}{} &\multicolumn{1}{c}{(s)}&\multicolumn{1}{c}{(m s$^{-1}$)}&&\multicolumn{1}{c}{(yrs)}
\\ \hline
HIRES & 45,000 & 270 $\times$ 4 exp & 11 & 1.2& 230 &12 \\
PFS & 80,000 & 235 $\times$ 4 exp & 40 &1.0 & 65 &5 \\
CHIRON & 95,000 & 120 $\times$ 3 exp & 300 & 1.6& 259 &2  \\
HARPS & 115,000 & 150 $\times$ 4 exp & 100& 0.4 & 437 &10 \\ 
\hline
\end{tabular}
\vspace{0.5cm}		
\\
\end{table*}

\subsection{HIRES Observations}

The full HIRES \citep{Vogt1994} dataset comprises 229 individual Doppler measurements with an observational baseline of almost twelve years, between November 22nd 2001 and August 25th 2013. These individual radial velocities have been binned nightly to produce 90 measurements. One outlier point with a velocity value more than 3-$\sigma$ away from the mean of the series has been rejected as it was acquired under poor weather conditions. 

HIRES uses the iodine cell method to deliver high precision radial velocities. The method employs a cell containing molecular gaseous iodine (I$_{2}$) that is mounted before the slit of the spectrograph so that the incoming starlight is imprinted with thousands of I$_{2}$ absorption lines, between $\sim$4800\AA{} and $\sim$6200\AA{} that are used for both very precise wavelength reference points and also in the determination of the instrumental point spread function (PSF). 

The HIRES spectrograph covers a wavelength range of 3700-8000\AA{}. For most of the observations the B5 Decker (0.86''$\times$3.5'') was used, delivering a spectral resolving power of $R\sim$45,000. The C2 Decker (0.86''$\times$14'', $R\sim$45,000) was also used for a smaller number of observations. I$_{2}$-free template observations were carried out with the B3 Decker (0.574''$\times$14'') at $R\sim$60,000. 

For the template observations we acquire multiple shots (typically 3) of the target star without I$_{2}$ with the narrow slit and we bracket these observations with the spectra of a bright, fast rotating B star observed through the I$_{2}$ cell. These I$_{2}$-free shots are then combined to create a high signal-to-noise, high resolution spectrum of the star that is later used for the computation of the radial velocities following the spectral synthesis procedure explained in \citet{Butler1996}, where the I$_{2}$ region is divided into $\sim$700 chunks of about 2\AA{} each to produce an independent measure of the wavelength, PSF, and Doppler shift.
This procedure is also carried out for PFS and CHIRON observations.

Exposure times varied with nightly weather conditions, but we obtained a formal mean\footnote{Weighted means using the radial velocity uncertainties as weights; $w_{i}= 1/\sigma_{i}$} uncertainty of $\sigma_{\rm BIN}$= 1.21 m s$^{-1}$ and $\sigma$=1.18 m s$^{-1}$ for the binned nightly and unbinned radial velocities, respectively, with this spectrograph. 

From individual HIRES spectra we have calculated the S-indices from the Ca \sc ii \rm H and K line cores (at 3968.47\AA{} and 3933.66\AA{}, respectively) following the prescription of \citet{Duncan1991} also described in \citet{Arriagada2011}. S-indices can be used for chromospheric activity analysis of the stars \citep{Arriagada2011, Boisse2011} since they are known to be correlated with spot activity on the surface of the star that can mimic planetary signals, or at best, introduce noise into the data. 

\subsection{PFS Observations}

Observations were carried out using PFS \citep{Crane2006, Crane2008, Crane2010} between October 18th 2011 and March 5th 2016. We obtained a total of 65 individual radial velocity measurements, translated into 19 binned velocities. PFS is also equipped with an I$_{2}$ cell for precise radial velocity measurements and it delivers a resolution of $R\sim$80,000 in the I$_{2}$ region when observing with the 0.5''$\times$2.5'' slit. I$_{2}$-free template observations were acquired with the 0.3''$\times$2.5'' slit at a resolution of $R\sim$127,000.

We routinely expose for a typical signal-to-noise ratio of $\sim$300 per spectral resolution element required to achieve a level of $\sim$1-2 m s$^{-1}$ radial velocity precision. For bright targets, such as HD26965, we take consecutive multiple exposures -usually 4 or 5- within a timespan of 5 minutes, to both average over the strongest stellar p-mode oscillations ($\sim$5 min for solar-type stars; \citealt{Leighton1962, EvansMichard1962,Ulrich1970}) and avoid saturation. For monitoring the stellar activity, S-indices were derived using individual spectra using the same approach described for HIRES.

We report a mean uncertainty of $\sigma$=0.97 m s$^{-1}$ from this instrument. Mean uncertainty for the nightly binned data is $\sigma_{\rm BIN}$= 0.98 m s$^{-1}$.

\subsection{CHIRON Observations}

All observations with the fiber-fed high-resolution echelle spectrograph CHIRON \citep{Tokovinin2013} were performed in service mode at $R\sim$95,000 using the `{\it Slit}' mode and 3$\times$1 pixel binning. CHIRON is installed in a thermally controlled space that allows the instrument to be stabilized to temperatures drifts of $\pm$ 2 K.The spectrograph covers a fixed wavelength range between 4150\AA{} and 8800\AA{} which, unfortunately, does not allow any measurement of calcium lines to monitor the chromospheric activity. CHI\-RON also employs an I$_{2}$ absorption cell for wavelength calibration. The CHIRON team provides reduced data corresponding to wavelength calibrated spectra \citep{Brewer2014}. We also acquired higher resolution I$_{2}$-free templates taken in `{\it Narrow}' mode at $R\sim$136,000 with the same pixel sampling as in `{\it Slit}' mode. Then we used our pipeline to compute the final Doppler shifts with a modified routine similar to the ones used in the PFS and HIRES reduction. 

In 2014 we started a high-cadence campaign using this instrument to monitor nearby bright FGK stars with V$\leq$6. When observing with CHIRON, we have found that the linearity regime for the CCD ends once the counts per pixel reach $\sim$30,000, so we have routinely exposed every target up to a maximum level of 25,000 counts to avoid reaching this non-linearity regime.
Since this target is a bright star, we use the same observational strategy that was used on both PFS and HIRES, meaning we take multiple short single exposures of the star that are combined into a single high-precision measurement.

Previous work by \citet{Jones2016} have shown precision of $\sim$5 m s$^{-1}$ using the high efficiency slicer mode to look for planets orbiting around giant stars, at a lower resolution of $R\sim$79,000 and for targets fainter than $V$=6.
Recent results by \citet{Zhao2018A} using the same observing mode we describe in this work have also shown consistent short-term (nightly) radial velocity precision on the $\sim$1 m s$^{-1}$ level for the very bright stars. They obtain a mean error of 1.1 m s$^{-1}$ and 1.2 m s$^{-1}$ for $\alpha$ Centauri A ($V$=-0.01) and B ($V$=1.13), respectively.

Results from our analysis give a mean radial velocity error for this bright star of $\sigma$=1.60 m s$^{-1}$ for the unbinned dataset consisting of 258 velocities taken between October 11th 2014 and January 15th 2016. The mean radial velocity error for the nightly binned velocities is $\sigma_{\rm BIN}$=1.62 m s$^{-1}$.

\subsection{HARPS Observations}

We used public data obtained with the HARPS spectrograph \citep{Mayor2003} available from the ESO HARPS archive\footnote{\url{http://archive.eso.org/wdb/wdb/eso/repro/form}}. All the data have been processed with the HARPS-Data Reduction Software (hereafter DRS) Version 3.5 pipeline which performs all the required reduction steps from bias, flat fielding and wavelength calibration of the high resolution spectra. HARPS is a pressure and temperature stabilized spectrograph that covers a wavelength range between 3800\AA{} and 6900\AA{} with a spectral resolving power of $R\sim$115,000. We note that HARPS does not make use of an I$_{2}$ cell for precise Doppler spectroscopy. Instead, exposures of a Thorium-Argon lamp are taken at the same time as each observation to get a precise wavelength reference for the science spectra (one spectrum on each of the two fibers that feed the instrument).
Radial velocities are a product of a post-reduction analysis consisting of cross-correlating each echelle order with a binary mask that is chosen depending on the spectral type of each star. This produces cross-correlation functions (CCF) for each order that are then combined to obtain a mean-weighted CCF.  This mean-weighted CCF is then used to generate the radial velocities. 
For HD26965 we found 483 useful public Doppler measurements between October 27th 2003 and December 5th 2013 available from ESO HARPS archive. The DRS pipeline and further post-reduction analysis produced 437 radial velocity measurements with a mean error of $\sigma$=0.43 m s$^{-1}$, and yielded a set of 65 binned radial velocities with a mean uncertainty of $\sigma_{\rm BIN}$=0.42 m s$^{-1}$. HARPS vacuum enclosure was opened in 2015 as part of an upgrade on the fibers. We refer to the pre-upgrade data as HARPS OLD. 
We include 82 post-upgrade HARPS velocities between September 9th 2015 and March 27th 2016. This post-upgrade data is labeled HARPS NEW.

For all the analyses, the unbinned data from each instrument is used and is treated separately with their corresponding independent velocity offset and noise (jitter) properties.

In the case of the spectrographs equipped with an I$_{2}$ cell (HIRES, PFS, CHIRON), the reported velocities are the weighted mean of the velocities of the individual chunks while the uncertainties correspond to the standard deviation of all the chunk velocities about that mean.
 For HARPS, where the observations were carried out using simultaneous Thorium exposures, the RV uncertainty is provided by the DRS and it is estimated directly from a Gaussian fit to the CCF \citep{Bouchy2001}.

{The 1,111 radial velocity measurements are shown from Table \ref{tab:firstset} to Table \ref{tab:lastset}.

\begin{figure*}\label{fig:pergram}
\centering
\vspace{-1.cm}
\includegraphics[scale=0.65]{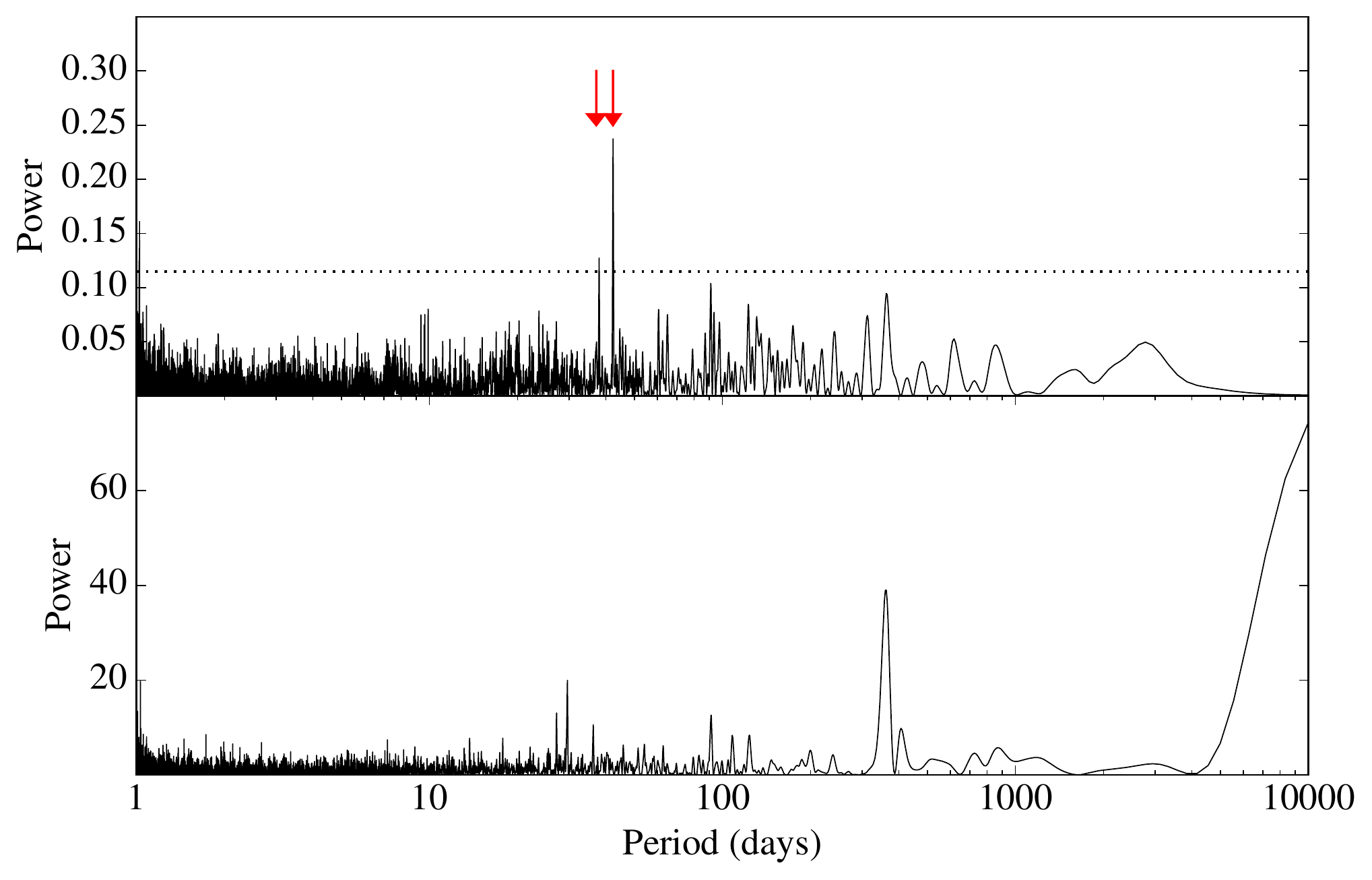}
\caption{{\it Top:} Generalized Lomb-Scargle periodogram of the unbinned combined velocities. The highest power is found at a period of 42.43 days. The vertical red arrows mark the position of the stellar rotation period and the period found in the time series. The dotted line shows the 0.1\% significance level, determined by 1,000 bootstrap resamplings. {\it Bottom:} Periodogram of sampling (window function) for the combined data.}
\end{figure*}

\section{Periodogram Analysis}\label{sec:pergram}

We started to examine the radial velocity data by using the traditional periodogram a\-na\-lysis approach to look for any periodicities embedded in the data. We used the generalized version \citep{Zechmeister2009} of the Lomb-Scargle periodogram (\citealt{Lomb1976, Scargle1982}, hereafter GLS) where we set up a minimum period of 1 day and a maximum period of 10,000 days for the search, with 80,000 trial periods evenly spaced in the frequency domain.

Figure \ref{fig:pergram} (top panel) shows the GLS periodogram of the combined radial velocities of HIRES, PFS, CHIRON and HARPS. The velocities have been mean subtracted. A maxima at 42 days (marked with a red arrow) clearly exceeds the power threshold of 0.1\% significance level. There are also two power maxima close to the 0.1\% significance threshold at $\sim$38 days and at $\sim$360 days. In the bottom panel of Figure \ref{fig:pergram} we show the periodogram of the sampling (window function) of the combined radial velocities. The secondary power spectrum peak at 360 days found in the periodogram of the velocities is also present here, and therefore can be attributed to the frequency of sampling. However, the peak at 38 days is not present and therefore further investigation is required to determine the origin of this possible signal, which we discuss below as being due to the rotation period of the star.
 
\section{Bayesian Analysis}\label{sec:bayes}

In addition to the traditional periodogram analysis we have performed a Bayesian analysis to search for periodic signals embedded in the data. 
We modeled the radial velocities of HD26965 following the statistical model defined in \citet{Tuomi2014} and also applied in \citet{JenkinsTuomi2014} where we include the following elements: 

\begin{itemize}[]
\item[] $1)$ A function describing a $k$-Keplerian planet model
\item[] $2)$ A linear trend term
\item[] $3)$ A red-noise model consisting of a $p$-th order moving average - MA($p$) - model with an exponential smoothing
\item[] $4)$ Linear correlations with the stellar activity indicators
\end{itemize}

We write the statistical model as follows
\begin{eqnarray}\label{eq:bayesmodel}
y_{i,j} = \gamma_{j} + \dot{\gamma}t_{i} + f_{k}(t_{i}) + \epsilon_{i,j} + \sum_{n=1}^{q_{j}} c_{n,j} \xi_{n,i,j} \nonumber\\
+\,\sum_{l=1}^{p}\phi_{j,l}\text{ exp }{\left\{ \frac{t_{i-l} - t_{i}}{\tau_{j}} \right\}} r_{i-l,j}
\end{eqnarray}

\noindent where $y_{i,j}$ corresponds to the observation at time $t_{i}$ for the $j$-th instrument, $\gamma_{j}$ is the velocity offset for the $j$-th dataset, $\dot{\gamma}$ is a linear trend term,  and $r_{i,j}$ denotes the residuals after subtracting the model from the measurement. The function $f_{k}$ is a superposition of $k$-Keplerian signals,

\begin{equation}\label{eq:keplerian}
f_{k} (t_{i}) =\sum_{m=1}^{k} K_{m} [\, \text{cos}( \omega_{m}+ \nu_{m}(t_{i})) + e_{m} \text{cos}(\omega_{m}) ]\, 
\end{equation}
where $K_{m}$ is the velocity semi-amplitude, $\omega_{m}$ is the longitude of pericenter, $\nu_{m}$ is the true anomaly and $e_{m}$ is the eccentricity. $\nu_{m}$ is also a function of the orbital period and the mean anomaly $M_{0,m}$. Hence, $f_{k}$ is fully described by $K_{m}$, $\omega_{m}$, $e_{m}$, $M_{0,m}$ and $P_{m}$ {\bf, $m\in\{1,...,k\}$}.\newline
The white noise term is denoted by the additive random variable $\epsilon_{i,j}$. We assume that there is an excess white noise in each data set with a variance of $\sigma_{j}$ such that
\begin{equation}
\epsilon_{i,j} \sim \mathcal{N}(0,\sigma^{2}_{i}+\sigma^{2}_{j})
\end{equation}

\noindent with $\sigma_{i}$ the uncertainty associated with the measurement $y_{i,j}$ and $\sigma_{j}$ is the excess white noise or jitter for the $j$-th dataset, that is treated as a free parameter in the model.

The remaining terms define the rest of the noise model, including the red-noise component: the first term with parameters $c_{n,j}$ describes the linear correlations with $q$ stellar activity indicators $\xi_{n,i,j}$ for the $n$-different instruments. The second term is the MA($p$) component with smoothing over a timescale $\tau_{j}=4$ days and $\phi_{j,l}$ with a value between -1 and 1 to quantify the correlation between measurements. The smoothing timescale is set to 4 days for simplicity \citep{Tuomi2013b}. We assume the noise is correlated in this timescale although with higher cadence smaller timescales would likely be more appropriate \citep{Tuomi2013b, Feng2016}.

\subsection{Posterior Samplings and Signal Detection}

To estimate the posterior probability of the parameters in the model given the observed data we use Bayes rule that states
\begin{equation}\label{eq:bayesrule}
P(\theta\, | \,y) = \displaystyle \frac{ P(y\,| \,\theta)\, P(\theta) }{\,\int P(y \,| \,\theta)\,P(\theta)\,  d\theta}
\end{equation}

\noindent where $ P(y\,| \,\theta)$ is the probability density of the measurements given the parameters (likelihood function) and $P(\theta)$ corresponds to the prior, i.e., what is known about a given parameter and its constraints before making the measurement. The denominator in equation \ref{eq:bayesrule} is a normalizing constant such that the posterior must integrate to unity over the parameter space.

In our model we chose the priors for the orbital and instrumental parameters as listed in Table \ref{table:priors}.

\begin{table}\label{table:priors}
\center
\caption{Prior selection for the parameters}
\begin{tabular}{lccc}
\hline\hline
\multicolumn{1}{l}{Parameter}& \multicolumn{1}{c}{Prior Type}&\multicolumn{1}{c}{Range}\\ 
\hline
Semi-amplitude&Uniform& $K\in [\,0,K_{\rm max} ]\,$ \\
Period& Jeffrey's&$P\in [\, 1, 2P_{\rm obs}]\,$\\
Eccentricity& $\mathcal{N}(0,\sigma_{e})$ &$e \in [0,1)$\\
Long. of Peric.& Uniform &$\omega \in [\, 0,2\pi]\,$\\
Mean Anomaly& Uniform &$M_{0} \in [\, 0,2\pi]\,$\\
Jitter& Uniform & $\sigma_{J} \in [0, K_{\rm max}] $\\
Smoothing time scale & Constant& $\tau_{j}$=4\\
\hline
\vspace{0.2cm}
\end{tabular}
\end{table}

In order to investigate the signal initially found with GLS periodogram of the combined radial velocities we use our Bayesian detection method where we sample the parameter space using the Delayed-Rejection Adaptive-Metropolis (DRAM) algorithm \citep{Haario2006} based on the Adaptive-Metropolis (AM) algorithm \citep{Haario2001},  applied in \citet{Tuomi2014a} and \citet{JenkinsTuomi2014}. DRAM and AM are both methods for improving the efficiency of the Metropolis-Hasting algorithm \citep{Metropolis1953, Hastings1970}. The idea behind using DRAM is that when the posterior of a parameter is multimodal, such as the orbital period in the case of Keplerian fits to radial velocity datasets, and a new state for the chain is rejected (see full details in \citealt{Tuomi2014}), a new proposed state is drawn centered on the last one.  Up to three rejections are allowed before that part of the posterior is finally discarded as a region of low probability.  This has the benefit of sampling more heavily the posterior phase space, at the cost of a longer run-time.

Tempered samplings are also performed when searching for signals. We include a $\beta$ parameter following \citet{Tuomi2014}, such as $\beta \in (0,1)$, meaning we use $P(\theta\, | \,y)^{\beta}$ instead of the standard posterior probability density, $P(\theta\, | \,y)$.  This way we can define the ``temperature'' of the chain simply as $T = 1/\beta$ and so a ``hot'' chain is defined when $T>1$ and a ``cold''  chain is where $T=1$. When $T>1$ the relative height of the maxima in the posterior probability density are decreased to prevent the chains from getting stuck in regions of high probability, allowing them to visit the entire period parameter space.  The typical length of a chain is set to be between $10^{6} - 10^{7}$ for the search run and $10^{6}$ for the initial burn-in period.

We performed a first run for a zero-planet model to determine the observational baseline, and the instrumental noise and stellar noise parameters for each set of radial velocities. We then searched for a signal in the radial velocity data considering a 1-planet model. The search was initially done by setting the temperature for the chain hot enough to let the chain explore the entire parameter space. This is especially helpful when the parameter space is highly multimodal. Our tolerance threshold for the acceptance rate is based on the optimal acceptance rate of the Metropolis-Hastings algorithm which is $\sim$0.234 \citep{Roberts1997}. A lower threshold for the chain to be accepted was set to 10\%, so hot chains with lower acceptance rates were discarded. From these runs we found a strong signal was present with a period of 42 days. The signal identified from the maximum of the posterior probability density distribution is shown in Figure \ref{fig:posterior}, left panel. 

We repeated this process by adding additional signals to the model, but we found no more statistically significant periods in the distribution of the posterior probability densities.
Finally, to constrain the detected signal, we performed parameter estimations via the AM algorithm by setting a cold chain ($\beta$=1) with the parameters initially set as a small ball around the parameters found previously by the hot chain run with DRAM.  This gave rise to the posterior histograms shown in Figure \ref{fig:param_dist}, where the period, amplitude, and minimum mass distributions show nice Gaussian forms centered on their respective values, and the eccentricity distribution is consistent with zero.
Table \ref{tab:system} summarizes the final set of values for the parameters from our analysis.
\begin{table}
\center
\caption{Solutions for HD26965. Final set of orbital and instrumental parameters. 1$\sigma$ errors.}
\label{tab:system}
\begin{tabular}{lcc}
\hline\hline
\multicolumn{1}{l}{Parameter}& \multicolumn{1}{c}{ \hspace{2cm}   }&\multicolumn{1}{c}{HD26965
~$b$} \\ \hline
$P$ (days) &       &42.364 $\pm$ 0.015  \\
$K$ (m s$^{-1}$) && 1.59 $\pm$ 0.15 \\
$e$ & &0.017$\pm$ 0.046 \\
$\omega$ (rad) &&  0.31 $\pm$ 1.93  \\
$M_{0}$ (rad) & & 4.92 $\pm$ 1.92  \\
$a$ (AU) & &0.215 $\pm$  0.008 \\
$m \sin i$ (M$_{\oplus}$) & & 6.91 $\pm$ 0.79  \\
$\gamma_{\rm PFS}$ (m s$^{-1}$)& & 0.13 $\pm$ 0.80 \\
$\gamma_{\rm HIRES}$ (m s$^{-1}$)& & 0.50 $\pm$ 0.57 \\
$\gamma_{\rm CHIRON}$ (m s$^{-1}$)& & 0.43 $\pm$ 0.79 \\
$\gamma_{\rm HARPS, old}$ (m s$^{-1}$)& & 0.17 $\pm$ 0.50 \\ 
$\gamma_{\rm HARPS,new}$ (m s$^{-1}$)& & 0.45 $\pm$ 0.78 \\ 
$\dot{\gamma}$ (m s$^{-1}\,$year$^{-1}$) & & -0.031 $\pm$ 0.037 \\\hline
$\sigma_{\rm PFS }$ (m s$^{-1}$)& & 1.54 $\pm$ 0.20  \\ 
$\sigma_{\rm HIRES}$ (m s$^{-1}$)& & 2.38 $\pm$ 0.15 \\ 
$\sigma_{\rm CHIRON}$ (m s$^{-1}$)& & 1.78 $\pm$ 0.15  \\ 
$\sigma_{\rm HARPS, old}$ (m s$^{-1}$)& &  1.11 $\pm$ 0.05 \\ 
$\sigma_{\rm HARPS,new}$ (m s$^{-1}$)& &  0.69 $\pm$ 0.07 \\ 
$\phi_{\rm PFS}$& & 0.82 $\pm$ 0.10 \\
$\phi_{\rm HIRES}$& & 0.61 $\pm$ 0.07 \\
$\phi_{\rm CHIRON}$& & 0.62 $\pm$ 0.06  \\
$\phi_{\rm HARPS, old}$ && 0.81 $\pm$ 0.04  \\ 
$\phi_{\rm HARPS,new}$ && 0.90 $\pm$ 0.08 \\ 
c$_{\rm S \, PFS}$ (m s$^{-1}$) && 61.1$\pm$ 14.0 \\
c$_{\rm S \, HIRES}$ (m s$^{-1}$) & & 53.1 $\pm$ 15.0 \\ 
c$_{\rm BIS \, HARPS, old}$ &&  0.086 $\pm$ 0.046 \\
c$_{\rm FWHM \, HARPS, old}$ && 1.8 $\pm$ 4.5  \\
c$_{\rm S \, HARPS, old}$ (m s$^{-1}$) && 1.6 $\pm$ 2.6\\
c$_{\rm H_{\alpha} \, HARPS, old}$ (m s$^{-1}$) & &-12.1 $\pm$ 8.9 \\
c$_{\rm He I \rm  \, HARPS, old}$ (m s$^{-1}$) & & -76 $\pm$ 34\\
c$_{\rm BIS \, HARPS,new}$ &&  0.27 $\pm$ 0.11 \\
c$_{\rm FWHM \, HARPS,new}$ && 0.089 $\pm$ 0.018  \\
c$_{\rm S \, HARPS,new}$ (m s$^{-1}$) && 105 $\pm$ 36\\
c$_{\rm H_{\alpha} \, HARPS,new}$ (m s$^{-1}$) & &76 $\pm$ 125 \\
c$_{\rm He I \rm \, HARPS,new}$ (m s$^{-1}$) & & 23 $\pm$ 141\\
\hline
\end{tabular}
\vspace{.5cm}
\end{table}
\subsection{Model Selection}
It is important to define a robust methodology that allows us to compare the results for two given models in order to address the statistical significance of one model with respect to the other.

The probability of a model $\mathcal{M}$, containing the best-fit parameters for the observed data $y$, is given by
\begin{equation}
P( \mathcal{M}\, | \,y) = \frac{P(y \, | \, \mathcal{M}) P(\mathcal{M})}{\sum^{k}_{i=1} P(y \, | \, \mathcal{M}) P(\mathcal{M})}
\end{equation}

In particular, we want to know if the model containing one planet is more probable than a zero-planet model and so on for the $k$-planet model with respect to a $k-1$-keplerian model. To solve this, we compute the probability of a given model by using the corresponding value of the Bayesian Information Criterion (BIC). A complete and detailed discussion can be found in \citet{TuomiJones2012} and \citet{Feng2016}. To compare the $\mathcal{M}_{k}$ model with a previous $\mathcal{M}_{k-1}$ model, we simply compute the logarithm of the Bayes factor, ln $B_{k,k-1}$, defined via

\begin{equation}\label{eq:bayesfactor}
{\rm ln}\, B_{k,k-1} = {\rm ln}\, P(y | \mathcal{M}_{k}) - {\rm ln}\, P(y | \mathcal{M}_{k-1})
\end{equation}

Furthermore, the model containing the best-fit parameters that support the signal has to fulfill the detection criteria described in \citet{Tuomi2012}. It must hold that

\begin{equation}
P(y | \mathcal{M}_{k}) = s P(y| \mathcal{M}_{k-1})
\end{equation}

\noindent where $s>10^{4}$. Hence using the Bayes factor defined in equation \ref{eq:bayesfactor}, we require that the $\mathcal{M}_{k}$ model describing the $k$-keplerian signal has to be more statistically probable than the $\mathcal{M}_{k-1}$ model associated with the $k-1$-keplerian signal. Following the conservative threshold from \citet{Tuomi2014}, we define that the evidence ratio should be

\begin{equation}
{\rm ln}\, B_{k,k-1} > 9.2
\end{equation} 
which translates posterior odds of 10,000:1 that the $k$ model is selected over the $k$-1 model, in order to satisfy our detection criteria.
Table \ref{tab:bfactors} shows the Bayes factors for $k=0,1,2$-planet models with and without activity correlation terms.

\begin{table}
\center
\caption{Logarithm of Bayes factors comparing a $k$=0, $k$=1 and $k$=2 Keplerian models with and without activity correlations}
\label{tab:bfactors}
\begin{tabular}{lcc}
\hline\hline

\multicolumn{1}{c}{\bf Bayes Factor}&\multicolumn{1}{c}{\bf Activity}&\multicolumn{1}{c}{\bf No Activity}\\
\multicolumn{1}{l}{\bf ln \boldmath$B_{k,k-1}$}&\multicolumn{1}{c}{\bf Correlations}&\multicolumn{1}{c}{\bf Correlations}\\ \hline 
ln $B_{1,0}$&43.38  & 35.44\\
ln $B_{2,1}$&0.61 & 2.96\\
\hline
\end{tabular}
\vspace{.5cm}
\end{table}

\begin{figure*}\label{fig:posterior}
\centering
\includegraphics[scale=0.34,angle=-90]{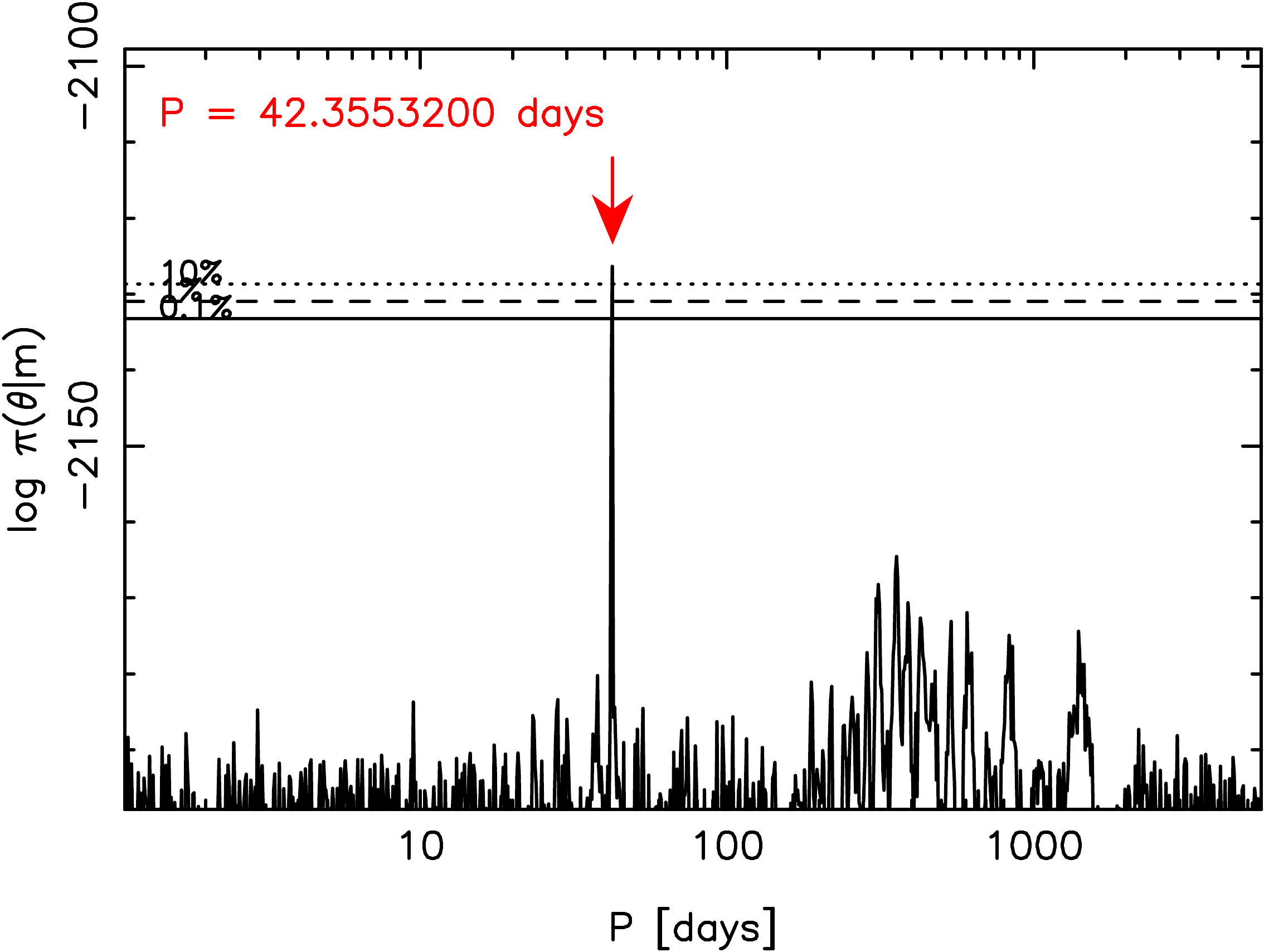}
\hspace{0.1cm}\vspace{1cm}\includegraphics[scale=0.33,angle=-90]{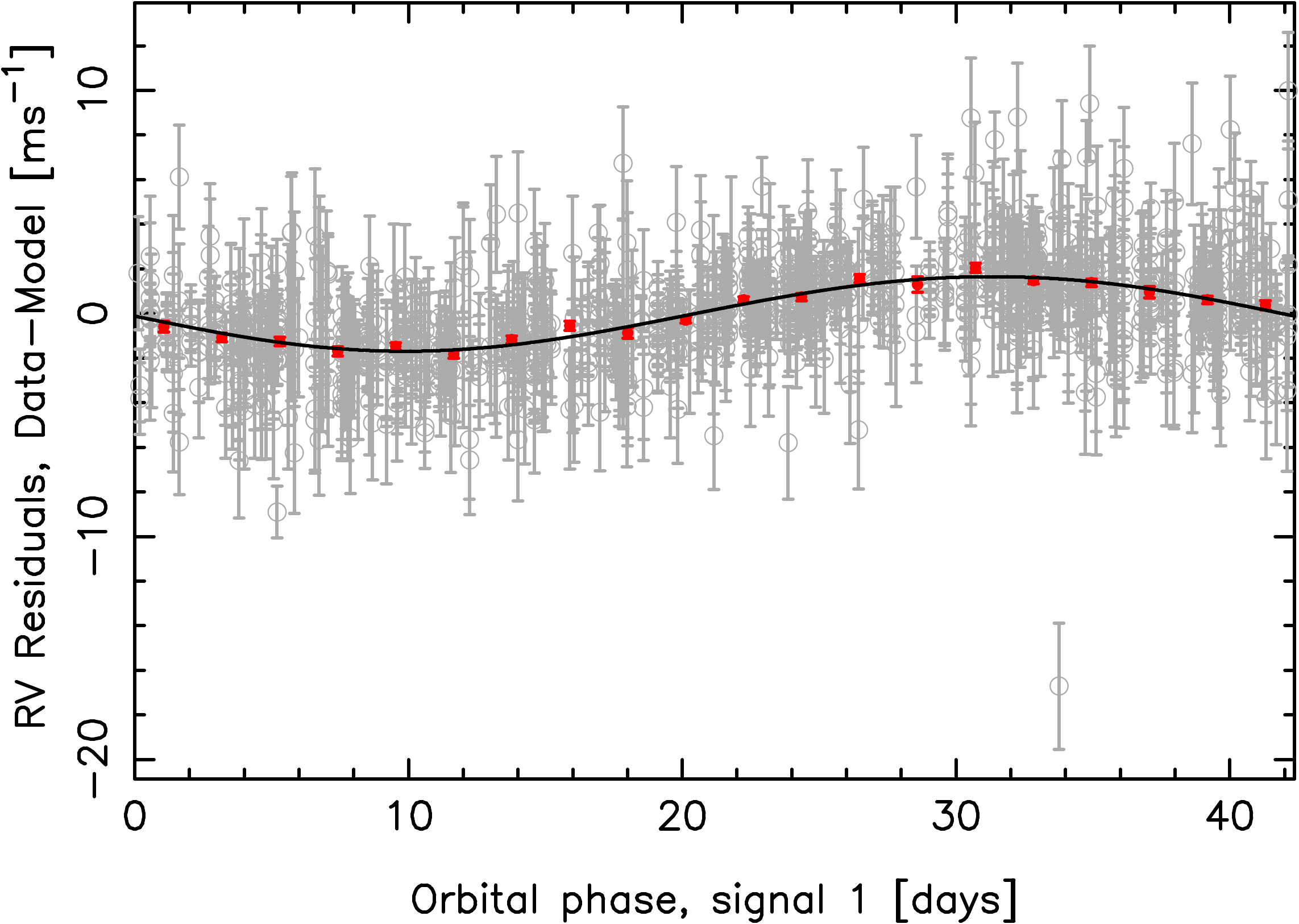}
\caption{{\it Left:} Posterior probability densities as output from our Bayesian code for a 1-planet model. {\it Right:} Phased-folded radial velocity curve.}
\end{figure*}

\begin{figure*}\label{fig:param_dist}
\centering
\includegraphics[scale=0.25,angle=-90]{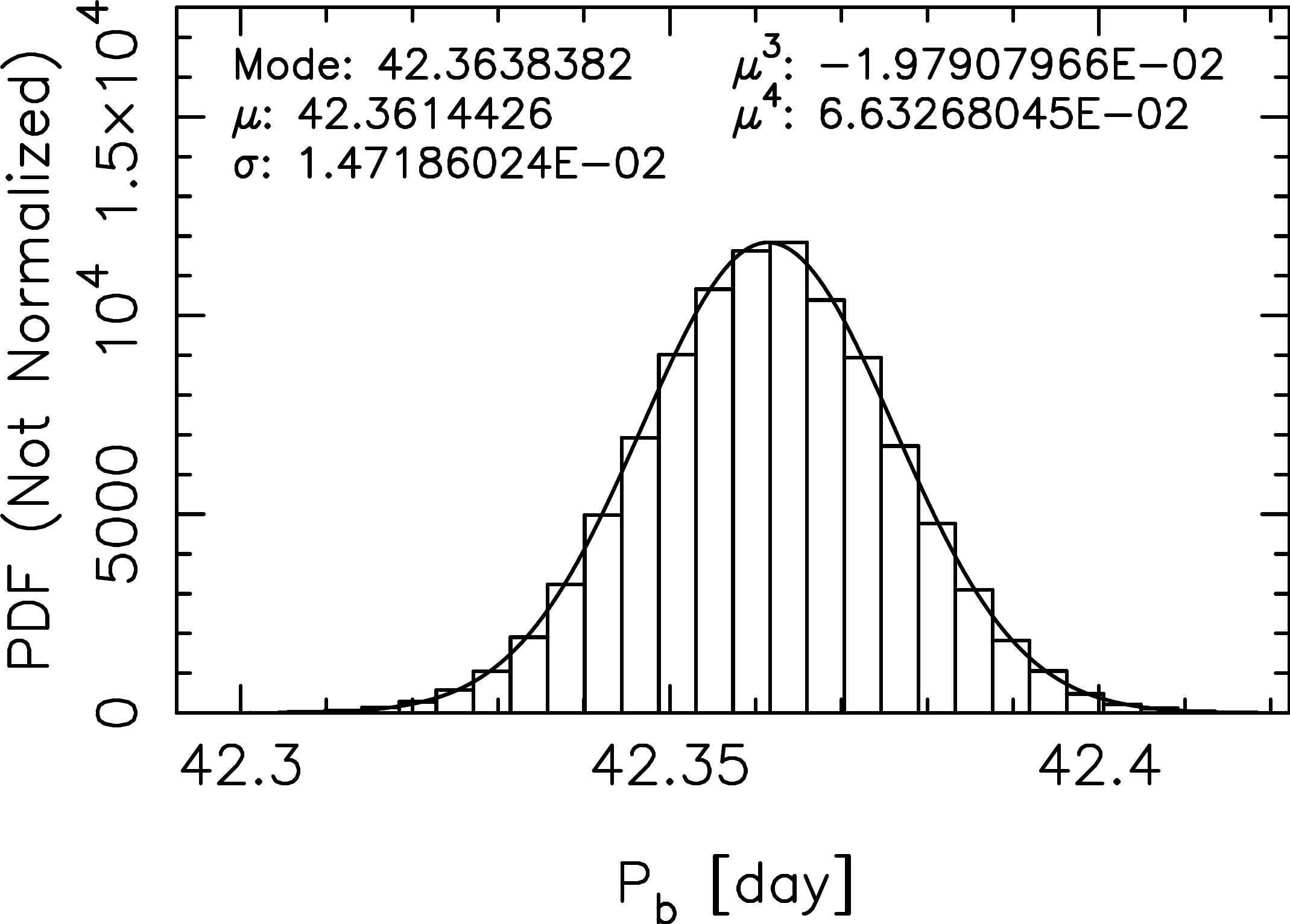}\includegraphics[scale=0.25,angle=-90]{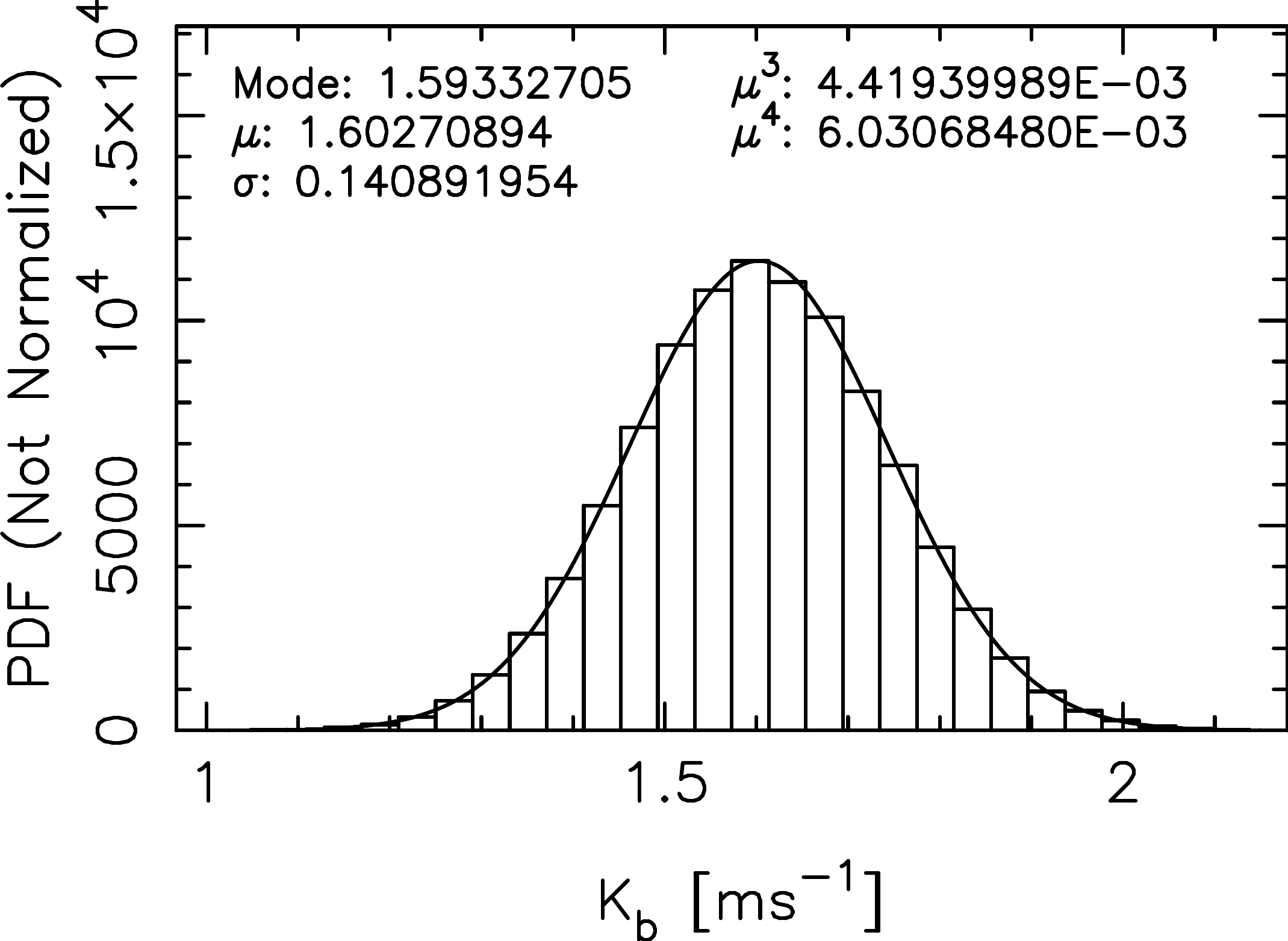}
\vspace{0.5cm}
\includegraphics[scale=0.25,angle=-90]{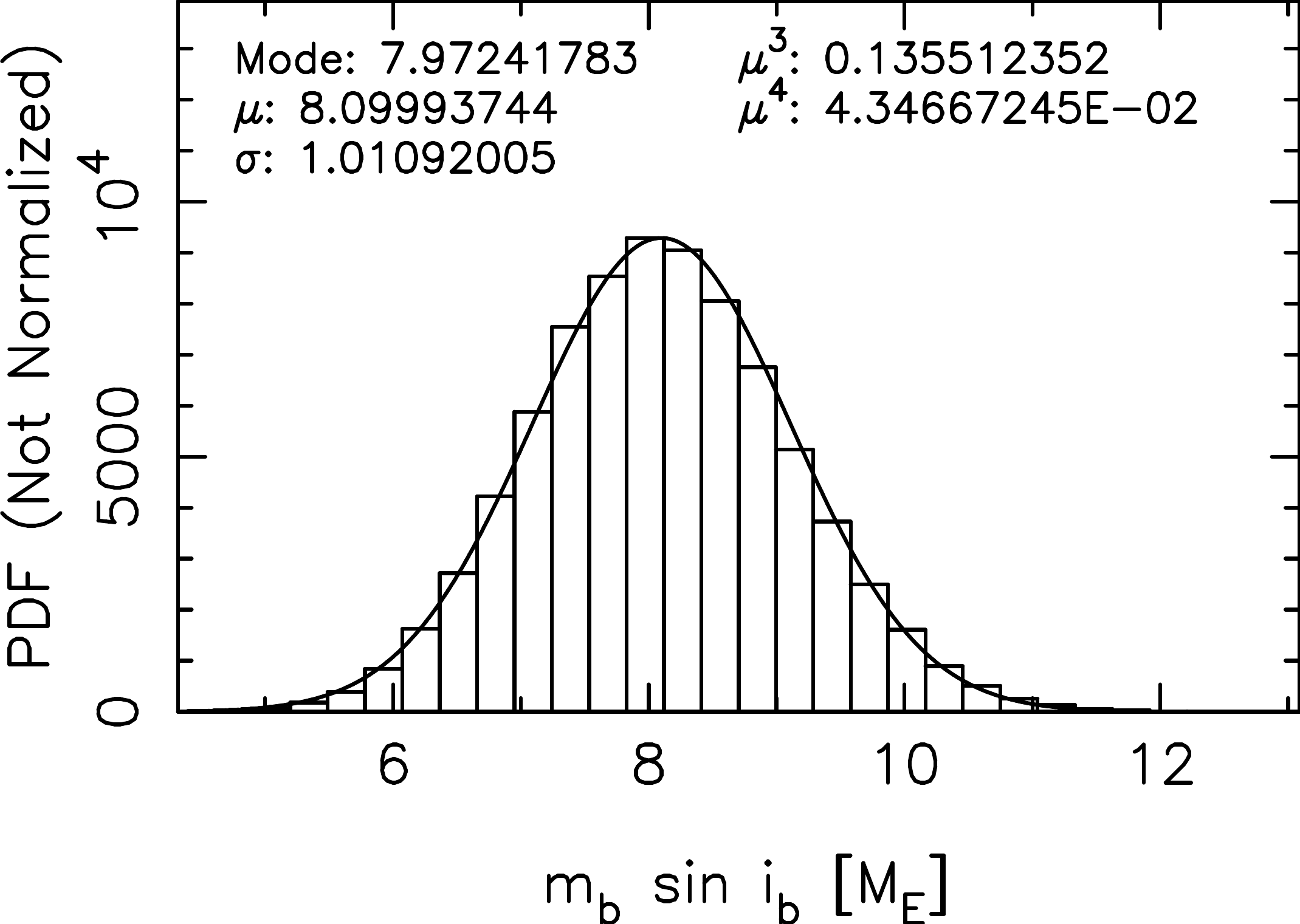}\includegraphics[scale=0.25,angle=-90]{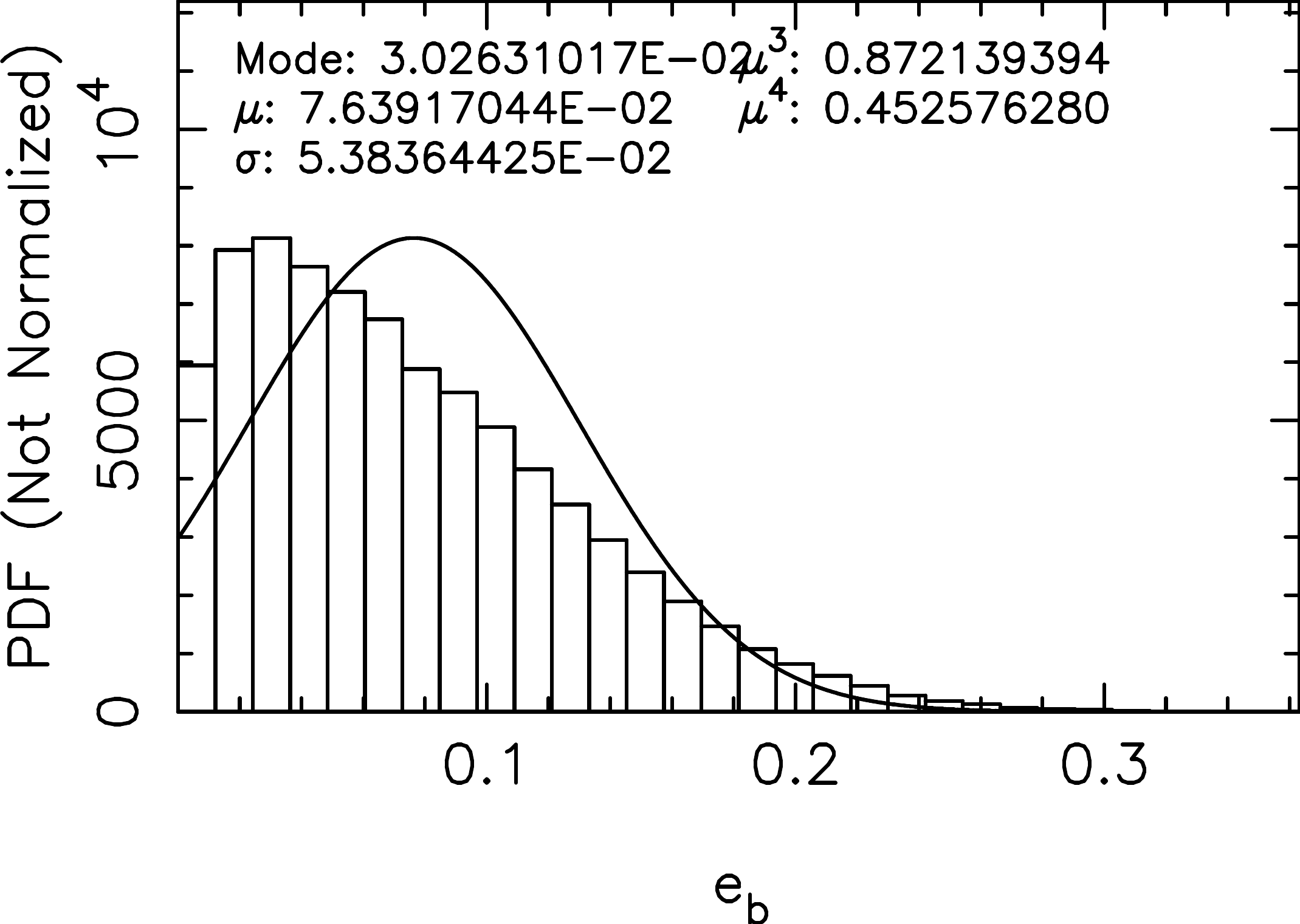}
\caption{{\it Left to Right, Top to Bottom:} Final distribution of period, semi-amplitude, minimum planetary mass and eccentricity resulting from a cold-chain Adaptive Metropolis run. The numbers at the top of each figure correspond to the mode, mean, variance, skewness and kurtosis, respectively. The solid line represents a Gaussian curve with same mean and variance.}
\end{figure*}

Figure \ref{fig:thresholds} shows the minimum mass detection thresholds for additional planets orbiting around HD26965. The green-filled area represents the liquid-water habitable zone estimated according to \citet{Kopparapu2013b, Kopparapu2013a}. The thresholds are calculated following the methods in \citet{Tuomi2014}. From this figure we can say that planets with minimum masses in excess of Neptune in the habitable zone can be ruled out meaning if there are HZ planets orbiting HD26965, they would likely be super-Earths or smaller. The red circle represents the planet candidate, barely, but significantly above the detection threshold.

\begin{figure}\label{fig:thresholds}
\centering
\includegraphics[scale=0.33,angle=-90]{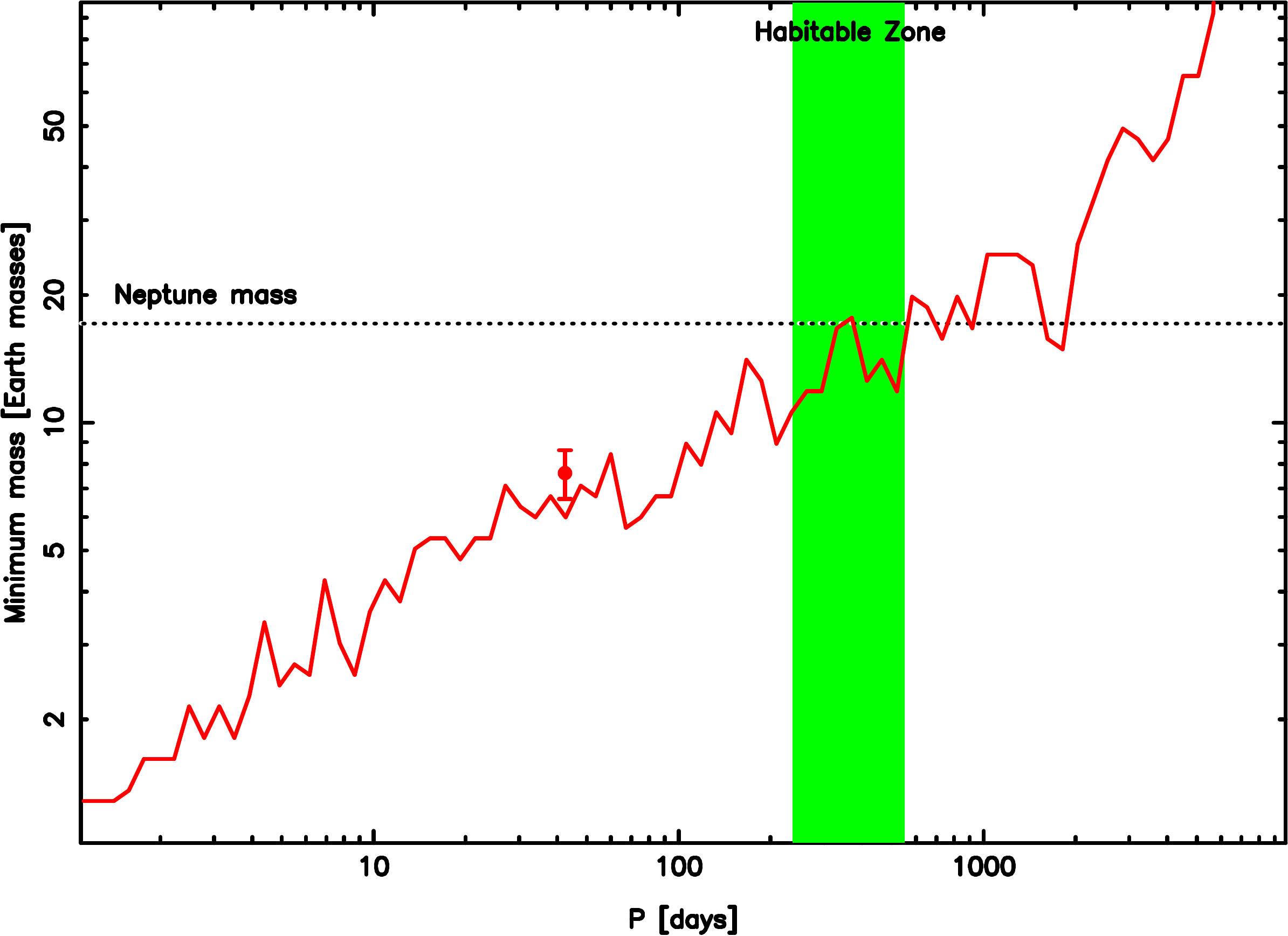}
\caption{Minimum mass detection thresholds for additional pla\-nets orbiting around HD26965 for periods between 1 and 10,000 days. The green-filled area highlights the habitable zone for this K dwarf.}
\end{figure}

\subsection{Signal injection}
\begin{figure}[ht]
\label{fig:injec_pergram}
\centering
\includegraphics[scale=0.6]{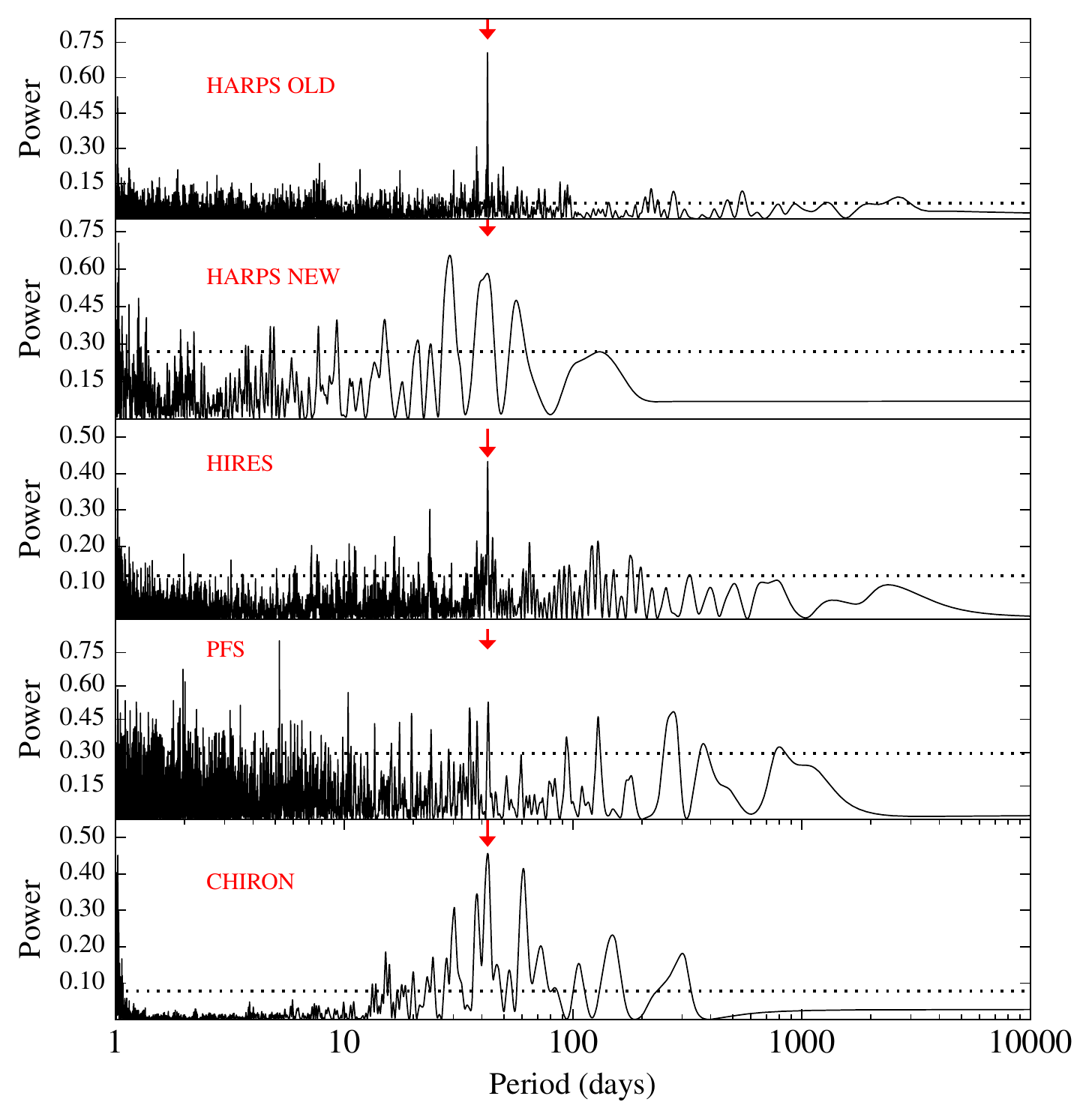}
\caption{{\it Top:} Generalized Lomb-Scargle periodogram for each dataset where a 1-planet model with the best-fit orbital parameters has been injected into the original measurements. Red arrows mark the candidate period found in the original time series. The dotted lines show the 0.1\% significance level, determined by 1,000 bootstrap resamplings.}
\end{figure}

As an additional test to investigate if the signal was supported for each instrument we performed a signal injection on the individual datasets. We use the best-fit parameters (i.e. $P$, $K$, $\omega$, $M_{0}$, e) of the putative signal by using a Keplerian function described in equation \ref{eq:keplerian}. The hypothesis is the following: if the 42-day signal is injected in a given dataset and we run our usual Bayesian analysis we should, in principle, easily detect it. If the signal is indeed present in the dataset we should recover $\sim$twice the best-fit radial velocity semi-amplitude as it has been boosted by the injection. On the other hand, if the recovered velocity semi-amplitude is significantly lower than our best-fit values, that would suggest the actual data is not supported by the instrument, or in other words, the precision of the instrument plus the current number of observations do not allow the signal to be detected. When boosting our signal, we recover the candidate period for  HARPS OLD, HIRES  and CHIRON datasets. For CHIRON, however, the expected peak at 42-days in the GLS is not unique, although it is above the 0.1\% significance level, as can be seen from the periodograms shown in Figure \ref{fig:injec_pergram}. In the case of PFS data, we did not recover the candidate period. Instead, we found a strong power at $\sim$5 days. This could be caused by the sparse sampling and lower number of observations available from this instrument.

\section{Stellar Activity and RV correlations}\label{sec:act_corr}
To investigate the nature of the detected 42 day signal, we perform a similar analysis as in \citet{Santos2014} on the activity indices available from each instrument. First we searched for periodicities present in the activity indices themselves, again using the GLS, and we show these results in Figure \ref{fig:act_per}. There are no statistically significant peaks associated with the 42 day signal we detect in the radial velocities. However, the periodogram of the HARPS S-indices shows an emerging peak at ~38 days, which is very close to the signal we detected in the radial velocities.  Interestingly, this was the period found for the rotation of the star from previous analysis of Ca \sc ii \rm lines \citet{Saar1997} which also agrees with the period inferred from ROSAT measurements (37.1 days; \citealt{Pizzolato2003}). 

Figure \ref{fig:correlations} shows the correlations between radial velocity and activity indicators for HARPS, PFS and HIRES. The combined S-indices we show have been mean subtracted and then combined together. We have computed the Pearson Rank test coefficients to determine the correlation between these quantities. Results are listed in Table \ref{tab:pearsonr} where we also list the uncertainties associated with each coefficient. To calculate these uncertainties we ran 10,000 bootstraps and created a distribution of $r$ coefficients for every activity index, where the standard deviation of the distribution gave us a measurement of the uncertainty on the coefficients. We note that the correlations are not significant within these uncertainties, given the standard statistical limits for claiming a weak ($r<$0.5), a moderate (0.5 $\le r \le$ 0.7), and a strong correlation ($r >$0.7), therefore we can conclude that the stellar activity indicators do not argue against a Doppler origin for the signal, yet the correlations indicate we must consider them in our full statistical model.  Indeed, the correlations suggest there is a weak impact of the stellar noise on the velocities, and we confirm this since the probability of our statistical model is higher when we include these correlations, compared to when we exclude them (see Table \ref{tab:bfactors}).

\begin{figure}\label{fig:act_per}
\centering
\includegraphics[scale=0.6]{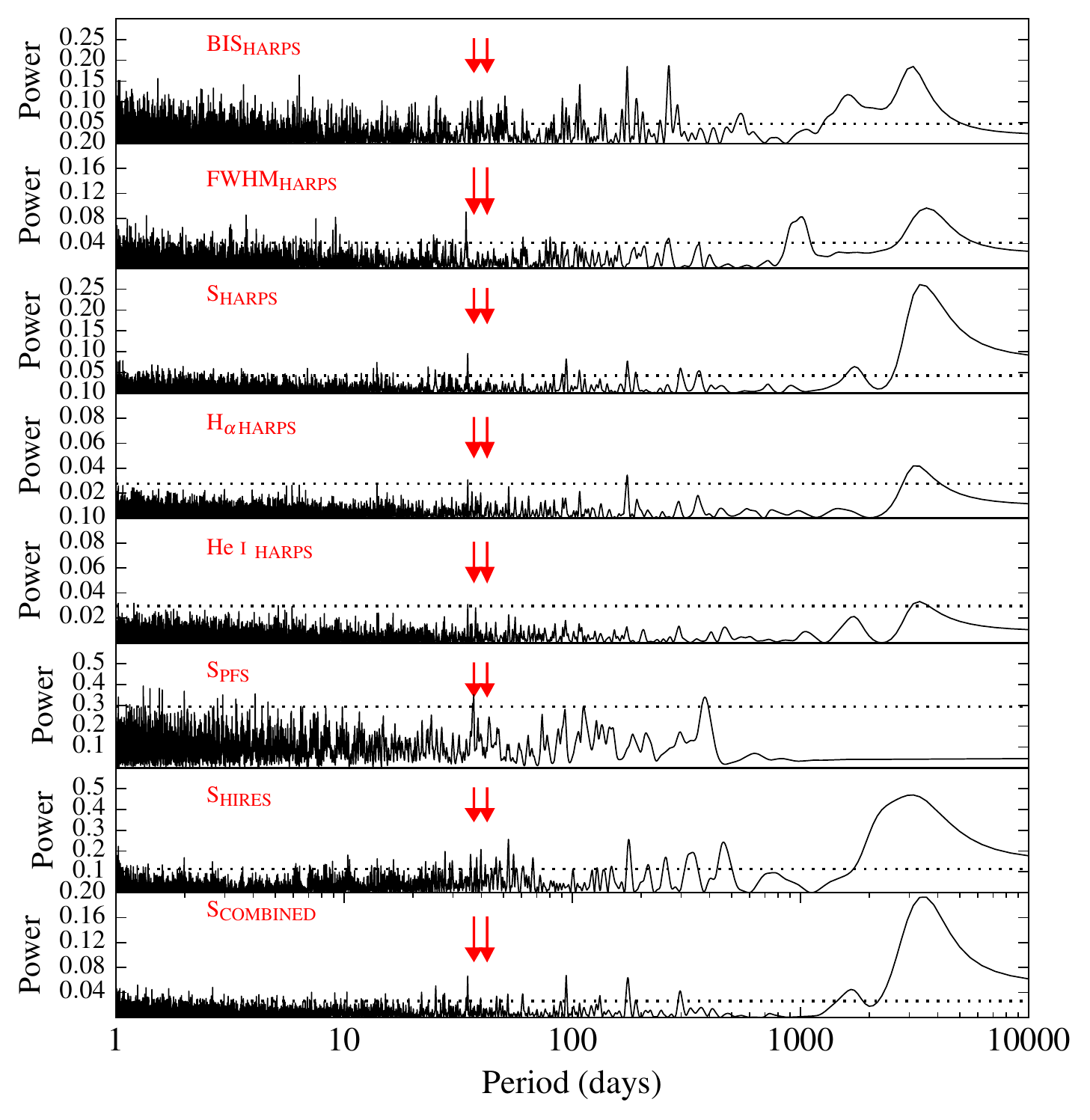}
\caption{Generalized Lomb-Scargle periodogram of the activity indicators available from the different spectrographs. From top to bottom: BIS$_{\rm HARPS}$, CCF FWHM$_{\rm HARPS}$, S$_{\rm HARPS}$, H$_{\alpha\, \rm HARPS}$, He \sc i\rm$_{\,\rm HARPS}$, S$_{\rm PFS}$, S$_{\rm HIRES}$ and S$_{\rm COMBINED}$. The arrows mark the position of the signal found at 42.37 days in the radial velocity series and the reported stellar rotation period of 37.1 days from \citet{Saar1997}. The dotted lines show the 0.1\% significance level, determined by 1,000 bootstrap resamplings. There are no statistically significant power in the activity indicators matching the radial velocity period.}
\end{figure}

\begin{figure*}\label{fig:correlations}
\centering
\includegraphics[scale=0.75]{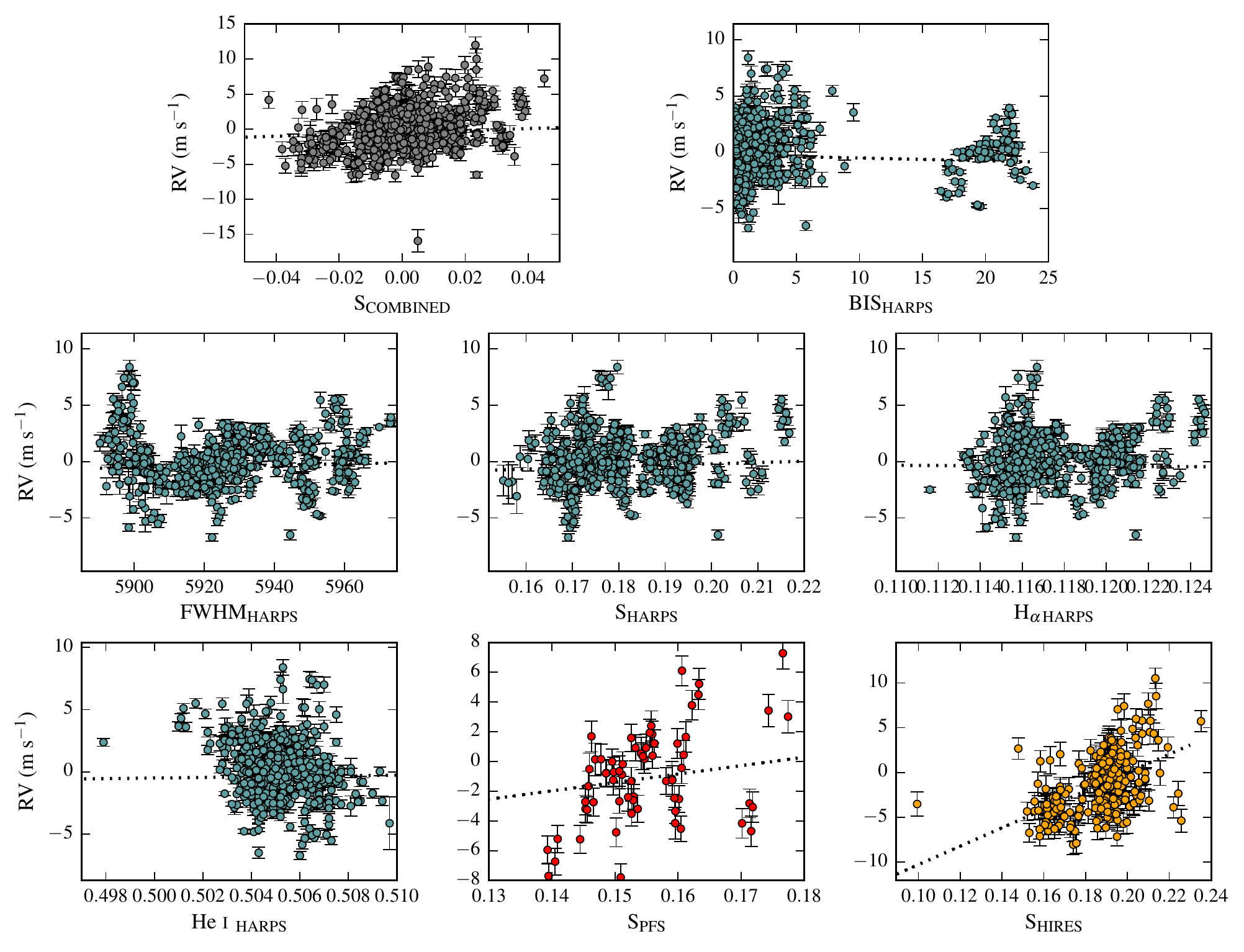}
\caption{Radial velocity correlations with respect to the seven different activity indicators: BIS$_{\rm HARPS}$, CCF FWHM$_{\rm HARPS}$, S$_{\rm HARPS}$, H$_{\alpha\, \rm HARPS}$, He \sc i \rm$_{\rm HARPS}$, S$_{\rm PFS}$, S$_{\rm HIRES}$. We also include the combined S-indices. The dotted lines mark the 1:1 relationships.}
\end{figure*}

\begin{table}
\center
\caption{Pearson Rank test coefficients. Correlation between activity indicators and radial velocities}
\label{tab:pearsonr}
\begin{tabular}{lcc}
\hline\hline
\multicolumn{1}{l}{Activity Indicator}& \multicolumn{1}{c}{\hspace{1cm}  }&\multicolumn{1}{c}{$r$}\\ \hline 
BIS$_{\rm HARPS}$& & -0.02 $\pm$ 0.04 \\
FWHM$_{\rm HARPS}$& & 0.74 $\pm$ 0.04 \\
S$_{\rm HARPS}$ && 0.14 $\pm$ 0.04 \\
H$_{\alpha \,\rm HARPS}$ && 0.05 $\pm$ 0.04 \\
He I$_{\,\rm HARPS}$& & -0.04 $\pm$ 0.04 \\
S$_{\rm PFS}$ && 0.48 $\pm$ 0.12 \\
S$_{\rm HIRES}$ && 0.44 $\pm$ 0.07 \\
S$_{\rm COMBINED}$ &&0.23 $\pm$ 0.03 \\
\hline
\end{tabular}
\vspace{.5cm}
\end{table}

\section{Testing variability and stability of the period and amplitude}\label{sec:variability}

Following a similar approach as in \citet{JenkinsTuomi2014}, we tested the variability and stability of the signal of our candidate. For this analysis we only considered the HIRES and HARPS OLD datasets, since both of them have a fairly continuous sampling of Doppler measurements along the $\sim$16 years of observational baseline. 
The measurements include a total of 662 unbinned velocities, and we chose JD$_{s}$=2454600 as the point to split the data, since this was close to the center of the time baseline of the observations and also produced a well balance between HARPS and HIRES data (i.e., not biased to an instrument in particular). The data prior to JD$_{s}$ contained 408 data points and the dataset after the split point contained 254 measurements.

We performed the Bayesian analysis on these 2 subsets of velocities independently, running cold chains to constrain the orbital parameters of a 1-Keplerian model. We found the signal is detected with values in agreement within uncertainties for the two baselines tested, as well as for the full data set described above. This shows us that the signal is not varying in time and thus the period and amplitude of our planetary candidate is stable over the tested observational baseline, another strong argument against an activity origin since activity processes should be quasi-static, varying over a few rotation periods of the star.

\begin{figure*}\label{fig:photometry}
\centering
\includegraphics[scale=0.56]{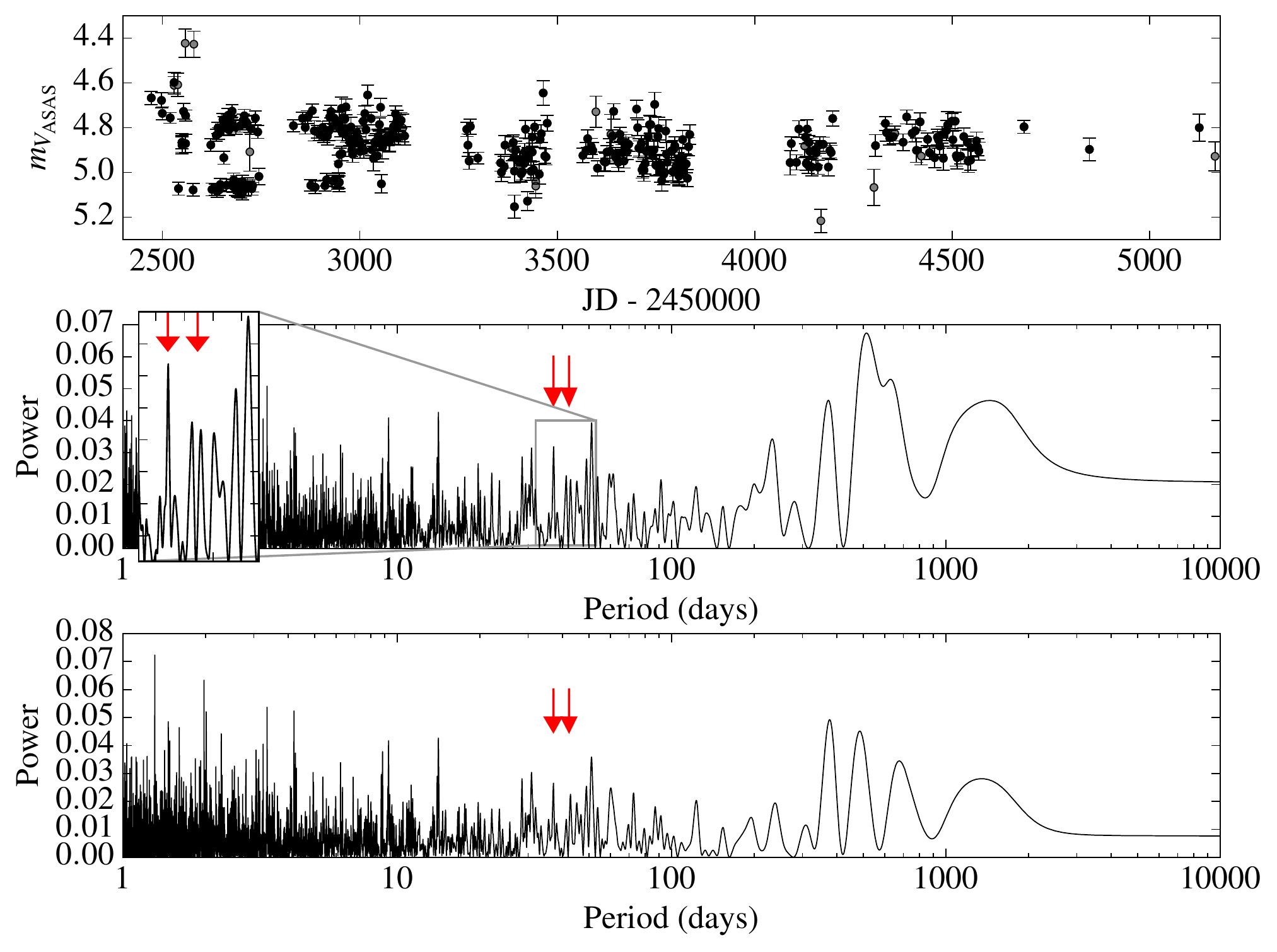}
\caption{{\it Top:} Photometric measurements from ASAS. Grey circles show the complete set of photometry while black-filled circles are those that meet the criteria as robust points described in the text. {\it Middle:} Generalized Lomb-Scargle periodogram of the ASAS photometry. Red arrows mark the position of the planetary candidate signal at 42.37 days and the stellar rotation period reported by \citet{Saar1997} of 38.7 days. No significant powers near the period - or an integer multiple of it - are found in the periodogram. The highest power is seen at 515 days. {\it Bottom:} Periodogram of the residuals after removing the 515 day period signal.}
\end{figure*}

\section{ASAS Photometry}\label{sec:photo}

To complement the analysis we gathered photometric data available from the All-Sky Automated 
Survey (ASAS) Catalog \citep{Asas1997} to investigate if any periodic signal could be seen in light curves, particularly the rotational period of the star. As mentioned above, the literature values were reported by \citet{Noyes1984} and \citet{Saar1997}, where they found $P_{\rm rot}\sim$37.10 days for this old star. We show the ASAS photometric measurements in Figure \ref{fig:photometry}. From the five different apertures available, we selected aperture 1 as its MAD\footnote{Median Absolute Deviation = median($\abs{x_{i} - {\rm median}(x)})$} value of 0.219 mag was the smallest. The mean uncertainty in the V-band photometry is $\sigma_{\rm ASAS}$= 0.036 mag. Grey circles correspond to the entire photometry set of 568 useful points acquired from 2000 to 2009. However, we excluded the data with poor quality (those not marked ``A'' or ``B'' in the catalog) and also those measurements that deviated more than 3-$\sigma$ with respect to the mean value of the time series. The highest quality data (316 points) are shown as black circles in the top panel of Figure \ref{fig:photometry}.
The bottom panel in Figure \ref{fig:photometry} again shows the GLS for the ASAS photometry. We sample the period space starting at a minimum period of 1 day and up to 10,000 days, performing 80,000 period samples. Considering just the data before JD$=$2452300 tends to favor peaks with higher power towards high frequencies (periods $\sim$1 day) but without any significant period (or an integer multiple) near the period associated with the 42 day signal of the reported planetary candidate. We also ran the periodogram analysis on the full photometric dataset with no significant periods found.  Following the relations in \citet{Hatzes2002}, we found that a filling factor of $f$=0.15 would be required to induce the RV amplitude of 1.6 m s$^{-1}$ of the signal found in the combined data. If we consider the spots on the surface of the star to be opaque, for the sake of simplicity, the ratio between the stellar flux and the flux considering spots covering 0.15\% of the surface of the star would be 0.9985. This means, the loss of light due to spots on the stellar surface can be translated into a $\Delta \, m$= 1.64 mmag. Given the precision of the ASAS photometry for this star, we conclude it is insufficient to be informative.

\section{Mount Wilson HK measurements}\label{sec:hk}

Given that we find some moderate correlations between the spectral activity indicators and the radial velocities, we supplemented our activity analysis by studying the original Ca \sc ii \rm H\&K data from the Mount Wilson Observatory HK Project \citep{Wilson1978}. The Project data are publicly available from the NSO archive\footnote{http://www.nso.edu/node/1335} and include more than 2,000 stars observed from 1966 to 1995.
There are 1,155 HK observations for HD26965 from JD=2439787.8 to 2449771.7. The processed data do not include associated uncertainties to the calibrated S-values. According to \citet{Duncan1991}, the uncertainties in the Mount Wilson S-values can be calculated using the weights, $W$, included in the data that are derived from the photon counts of the measurements. The uncertainty in the S-index measurements is simply defined as $\sigma_{S}= S \,(\sqrt{W})^{-1}$.  We applied this formula to the reported weights to provide proper uncertainties for the measurements of this star.  All HK values for HD26965 can be found in Table \ref{tab:hk_data}.

\begin{figure}\label{fig:hk}
\centering
\includegraphics[scale=0.57]{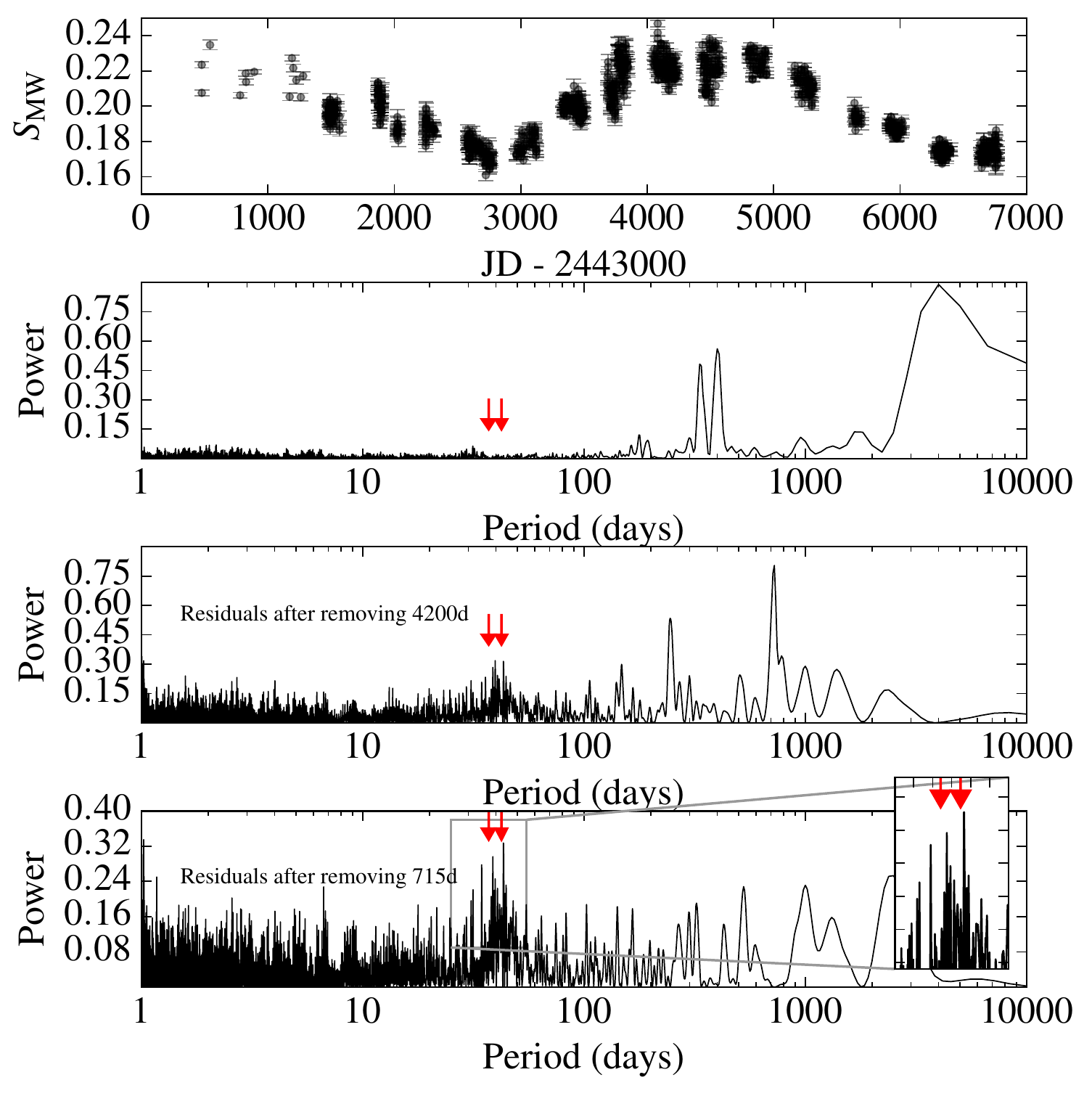}
\caption{{\it Top:} Mount Wilson S-values for HD26965. Second, Third \& Bottom Panels: Periodogram of the Mount Wilson S-values, periodogram of residuals after removing $\sim$4100 days and periodogram of residuals after removing $\sim$715 days, respectively. Red and black dashed lines mark the position of the possible rotation periods of $\sim$37 and $43$ days, respectively. }
\end{figure}

We proceeded to run the same periodogram analysis as for the radial velocities and the photometric time-series from ASAS.
A clear long-term variability of $\sim$4,100 days can be identified (see top and second top panels in Figure \ref{fig:hk}), providing evidence for a long-period magnetic activity cycle, similar to the long-period solar cycle. After removing this signal by modeling it with a sinusoidal function, a second period of $\sim$715 days is found in the periodogram (second bottom panel in the figure), likely representing another, shorter period magnetic cycle. Finally, the bottom panel in the figure shows the periodogram of the Mount Wilson S-values after removing the 4,100 and 715 day period signals from the data. In this residual periodogram a signal of  $\sim$42.3 days remains in the data. This is most likely the value reported in \citet{Baliunas1996}.
The peak is clearly not unique, casting some doubt on its reality, but given that it matches the detected signal in our radial velocity data sets, and rotation periods are known to be quasi-period due to differing spot patterns, changing stellar activity levels, and differential rotation, we must entertain the real possibility that this is actually the rotational period of the star, and not the 38 day period that we found in the measured spectroscopic activity indices.  If this is the reality, then HD26965 represents a case where most of the current suite of tests that we employ to detect planets using radial velocity analyses, fail to remove the noise introduced by the rotation of the star, meaning that we now require better methods to be employed on stars where there are clear correlations between the radial velocities and various activity indicators if we want to detect planets that induce amplitudes at the $\sim1$ m s$^{-1}$ level.

\section{Conclusions}\label{sec:conc}
Through the application of generalized Lomb-Scargle periodograms and tempered MCMC samplings, we conclude that there is a strong periodic signal in the radial velocities of the quiescent and slow rotating K dwarf HD26965. If interpreted as the Doppler signal induced on the star by an orbiting planet, our best solution explains these variations by the presence of a low-mass, super-Earth planetary candidate that has a minimum mass of 6.92$\pm$0.79 M$_{\oplus}$ orbiting the host star with a period of 42.364$\pm$0.015 days and at a distance of 0.215$\pm$0.008 AU.

Analysis of V-band photometry from ASAS does not show any significant periodic signal. However, since the amplitude of the signal is small, the precision of the data is not sufficient to detect the signal within the noise of the photometry.

Analysis of the stellar activity indicators does not show statistical evidence supporting a chromospheric origin for the periodic variations in the radial velocities, although we have found correlations between the radial velocities and the activity indices from the different spectrographs. However, when we analyze the independently acquired chromospheric calcium S-indices from the Mt. Wilson HK project, and after removing two long period activity cycles, we find evidence for the rotation of the star closely matching the period of the radial velocity detected signal.  

Regarding this last point, we note that although it is important to properly include activity correlations into any global model of radial velocity data, which when done for this data set we find a higher statistical probability for the given Keplerian model supporting the planetary signal, if there are statistically significant correlations between activity indicators and the velocity measurements, then additional external activity indicators should be acquired, where possible. Also, moving away from linear correlation models between current activity indices and the radial velocities may be necessary, particularly if the data suggests more complex models, such as quadratics, might be favored. In any case it is clear that the inclusion of multiple sources of external data that also rule out possible magnetic cycles and rotation periods as the source of any radial velocity signal, can help to maintain the lowest false-positive rate for any given Doppler survey. 

In summary, despite all the evidence favoring a Doppler signal present in this radial velocity data set, the methods described in this paper do not seem to be able to disentangle weak planetary signals from residual photospheric noise, at least when the orbital periods are close to the rotation period of the star and there are correlations present between the velocities and the measured activity indicators.

\acknowledgments
MRD acknowledges the support of CONICYT-PFCHA/Doctorado Nacional-21140646, Chile. JSJ acknowledges support by Fondecyt grant 1161218 and partial support by CATA-Basal (PB06, CONICYT). MGS is supported by CONICYT-PFCHA/Doctorado Nacional-21141037, Chile. The authors wish to recognize and acknowledge the very significant cultural role and reverence that the summit of Mauna Kea has always had within the indigenous Hawaiian community. We are most fortunate to have the opportunity to conduct observations from this mountain. This research has made use of the SIMBAD database, operated at CDS, Strasbourg, France.
\newpage 

\begin{table*}
\center
\caption{HIRES Radial Velocities of HD26965}\label{tab:firstset}
\label{tab:hiresrv}
\begin{tabular}{lccc}
\hline\hline
\multicolumn{1}{l}{BJD}& \multicolumn{1}{c}{RV}&\multicolumn{1}{c}{$\sigma$ RV}& \multicolumn{1}{c}{S}\\ 
\multicolumn{1}{l}{}& \multicolumn{1}{c}{(m s$^{-1}$)}&\multicolumn{1}{c}{(m s$^{-1}$)}& \multicolumn{1}{c}{(dex)}\\ \hline 
2452235.83300& -1.415& 1.2524& 0.1792\\
2452236.85549& -1.805& 1.3973& 0.1887\\
2452237.89810& 2.040& 1.3150 &0.1679\\
2452307.73757& -2.992& 1.4854& 0.1785\\
2452536.99956& -1.836& 1.4599& 0.1609\\
2452601.99297& -4.451& 1.2644& 0.1812\\
2452856.13402& 1.361& 1.5298& 0.1630\\
2452856.13536& -3.705& 1.3752& 0.1605\\
...&...&...&...\\
\hline
\end{tabular}
\tablecomments{This table is published in its entirety in the machine-readable format. A portion is shown here for guidance regarding its form and content.}
\end{table*}

\begin{table*}
\center
\caption{PFS Radial Velocities of HD26965}
\label{tab:pfsrv}
\begin{tabular}{lccc}
\hline\hline
\multicolumn{1}{l}{BJD}& \multicolumn{1}{c}{RV}&\multicolumn{1}{c}{$\sigma$ RV}& \multicolumn{1}{c}{S}\\ 
\multicolumn{1}{l}{}& \multicolumn{1}{c}{(m s$^{-1}$)}&\multicolumn{1}{c}{(m s$^{-1}$)}& \multicolumn{1}{c}{(dex)}\\ \hline 
2455852.81626& 0.0445& 1.1825& 0.1541\\
2455852.81753& -2.5663& 1.1740& 0.1590\\
2455852.81876&-4.0511& 1.0451& 0.1595\\
2455852.82000& -2.2046& 1.1274& 0.1606\\
2456175.89728& -0.5003& 0.9877& 0.1511\\
2456285.67699& -3.6936& 0.8562& 0.1526\\
2456285.67840& -2.8543& 0.8220& 0.1530\\
2456285.67978& -3.6332& 0.8654& 0.1529\\
... &...&...&...\\
\hline
\end{tabular}
\tablecomments{This table is published in its entirety in the machine-readable format. A portion is shown here for guidance regarding its form and content.}
\end{table*}
\vspace{-0.5cm}
\begin{table*}
\center
\caption{CHIRON Radial Velocities of HD26965}
\label{tab:chironrv}
\begin{tabular}{lcc}
\hline\hline
\multicolumn{1}{l}{BJD}& \multicolumn{1}{c}{RV}&\multicolumn{1}{c}{$\sigma$ RV}\\ 
\multicolumn{1}{l}{}& \multicolumn{1}{c}{(m s$^{-1}$)}&\multicolumn{1}{c}{(m s$^{-1}$)}\\ \hline 
2456941.80711& 0.7180& 1.5175\\ 
2456941.81054& 1.9459& 1.4766\\ 
2456941.81463& 0.9758& 1.6280\\ 
2456942.79556& 2.0399& 1.4729\\ 
2456942.79920& 0.6670& 1.5552\\ 
2456942.80289& 0.7219& 1.4767\\ 
2456943.75367& 1.8081& 1.5570\\
2456943.75739& 1.0934& 1.7136\\
...&...&...\\
\hline
\end{tabular}
\tablecomments{This table is published in its entirety in the machine-readable format. A portion is shown here for guidance regarding its form and content.}
\end{table*}
\vspace{-0.5cm}

\begin{table*}
\centering
\caption{HARPS OLD Radial Velocities of HD26965}\label{tab:lastset}
\begin{tabular}{lccccccc}
\hline\hline
\multicolumn{1}{l}{BJD}& \multicolumn{1}{c}{RV}&\multicolumn{1}{c}{$\sigma$ RV}& \multicolumn{1}{c}{BIS}& \multicolumn{1}{c}{FWHM}& \multicolumn{1}{c}{S}& \multicolumn{1}{c}{H$_{\alpha}$}& \multicolumn{1}{c}{He \sc i}\\ 
\multicolumn{1}{l}{}& \multicolumn{1}{c}{(m s$^{-1}$)}&\multicolumn{1}{c}{(m s$^{-1}$)}& \multicolumn{1}{c}{(m s$^{-1}$)}& \multicolumn{1}{c}{(m s$^{-1}$)}& \multicolumn{1}{c}{(dex)}& \multicolumn{1}{c}{(dex)}&\multicolumn{1}{c}{(dex)}\\ \hline
2452939.80613& -0.434& 0.519&1.254& 5896.904& 0.1753& 0.1161& 0.5067\\
2452939.80685& 0.231& 0.512&1.279& 5897.505& 0.1742& 0.1163& 0.507\\
2452939.80756& -0.367& 0.572& 1.387& 5901.748& 0.1747& 0.1163& 0.505\\
2452939.80827& 0.176& 0.524& 1.720& 5898.172& 0.1759& 0.1153& 0.5061\\
2452939.80899& 2.305& 0.954& 3.537& 5899.064& 0.1726& 0.1152& 0.5039\\
2452939.80969& 1.191& 0.612& 1.192& 5899.194& 0.1722& 0.1163& 0.5075\\
2452940.76906& -3.630& 0.378& 0.184& 5898.594& 0.175& 0.1157& 0.5076\\
2452945.76432& -3.375& 0.353& 2.281& 5896.436& 0.171& 0.1153& 0.5067\\
...&...&...&...&...&...&...&...\\
\hline
\end{tabular}
\tablecomments{This table is published in its entirety in the machine-readable format. A portion is shown here for guidance regarding its form and content.}
\end{table*}

\begin{table*}
\centering
\caption{HARPS NEW Radial Velocities of HD26965}\label{tab:lastset}
\begin{tabular}{lccccccc}
\hline\hline
\multicolumn{1}{l}{BJD}& \multicolumn{1}{c}{RV}&\multicolumn{1}{c}{$\sigma$ RV}& \multicolumn{1}{c}{BIS}& \multicolumn{1}{c}{FWHM}& \multicolumn{1}{c}{S}& \multicolumn{1}{c}{H$_{\alpha}$}& \multicolumn{1}{c}{He \sc i}\\ 
\multicolumn{1}{l}{}& \multicolumn{1}{c}{(m s$^{-1}$)}&\multicolumn{1}{c}{(m s$^{-1}$)}& \multicolumn{1}{c}{(m s$^{-1}$)}& \multicolumn{1}{c}{(m s$^{-1}$)}& \multicolumn{1}{c}{(dex)}& \multicolumn{1}{c}{(dex)}&\multicolumn{1}{c}{(dex)}\\ \hline
2457274.86039       & -0.1517   & 0.1828   &18.5315 &5951.1924   & 0.1777 &   0.1170    &0.5055\\
2457274.86304       &-0.0462    &0.1852   &18.4779 &5951.7017   & 0.1768  &  0.1171    &0.5059\\
2457274.86566       & 0.4185    &0.2041   &18.6366 &5951.8066   & 0.1777   & 0.1174    &0.5059\\
2457274.86845       &0.3169    &0.2026   &18.1923 &5951.7324   & 0.1778  &  0.1173   & 0.5051\\
2457277.84809       & 0.4107    &0.2189   &18.9974 &5950.3081  &  0.1763   & 0.1179  &  0.5056\\
2457277.85019       &  0.1997    &0.2209  & 18.6706 &5950.1743 &   0.1777    &0.1176 &   0.5049\\
...&...&...&...&...&...&...&...\\
\hline
\end{tabular}
\tablecomments{This table is published in its entirety in the machine-readable format. A portion is shown here for guidance regarding its form and content.}
\end{table*}

\begin{table*}
\center
\caption{ASAS Photometry of HD26965}\label{tab:asas_data}
\begin{tabular}{lcccccccccccc}
\hline\hline
\multicolumn{1}{l}{HJD -2450000} & \multicolumn{1}{c}{MAG 4}&\multicolumn{1}{c}{MAG 0}& \multicolumn{1}{c}{MAG 1}& \multicolumn{1}{c}{MAG 2}& \multicolumn{1}{c}{MAG 3}& \multicolumn{1}{c}{MER 4}& \multicolumn{1}{c}{MER 0}& \multicolumn{1}{c}{MER 1}&\multicolumn{1}{c}{MER 2}& \multicolumn{1}{c}{MER 3}& \multicolumn{1}{c}{GRADE}& \multicolumn{1}{c}{FRAME}\\ \hline
1953.56936 & 29.999 & 29.999 & 29.999& 29.999& 29.999 &   0.028 &0.063& 0.050& 0.034& 0.029 & C &9642 \\
2172.77457 &  5.375 & 5.451  &5.324&  5.361 & 5.374  &  0.033 &0.054& 0.035 &0.027& 0.029  &A &32848 \\
2206.76432 &29.999 & 29.999 &29.999& 29.999 &29.999  &  0.030 &0.084& 0.057 &0.037& 0.031  &C &37541 \\
2227.69007 & 5.274 & 4.831  &4.866 & 5.055  &5.187   & 0.030 &0.046 &0.041 &0.031 &0.032  &A &39837 \\
2230.68849 & 5.235 & 4.866  &4.942 & 5.091  &5.179   & 0.038 &0.107 &0.090 &0.063 &0.050  &D &40329 \\
2234.67705 & 5.263 & 4.766  &4.821 & 5.023  &5.175   & 0.032 &0.046 &0.040 &0.032 &0.035  &A &40805\\
2236.67316 &29.999 & 29.999 &29.999& 29.999 &29.999  &  0.034& 0.046& 0.039 &0.031& 0.036 & C &41118 \\
2501.90399 &29.999 & 29.999 &29.999& 29.999 &29.999  &  0.033& 0.044& 0.044 &0.035& 0.038 & C &16126 \\
2529.79004 & 4.598 & 3.968  &4.145 & 4.378  &4.523   & 0.044 &0.060 &0.052 &0.043 &0.046  &B &19144 \\
2549.77699 & 4.867 & 5.224  &4.976 & 4.923  &4.875   & 0.028 &0.060 &0.056 &0.040 &0.033  &B &20837 \\
2553.76674 & 4.727 & 4.377  &4.445 & 4.585  &4.681   & 0.035 &0.049 &0.052 &0.043 &0.041  &B &21379 \\
2558.77118 &29.999 & 29.999 &29.999& 29.999 &29.999  &  0.028& 0.052& 0.039& 0.028& 0.027 & C& 22103 \\
2655.59412 & 4.934 & 6.289  &5.861 & 5.440  &5.139  &  0.029 &0.036 &0.037 &0.027& 0.029  &A &35213 \\
2954.75032 & 4.916 & 5.227  &5.076 & 4.981  &4.927 &   0.032 &0.032 &0.037 &0.028& 0.029  &A &79986 \\
... & ... & ... & ... & ... & ... & ... & ... & ... & ... & ... & ... & ...  \\ \hline
\end{tabular}
\tablecomments{This table is published in its entirety in the machine-readable format. A portion is shown here for guidance regarding its form and content.}
\end{table*}

\begin{table*}
\center
\caption{Mount Wilson HK Project measurements of HD26965}\label{tab:hk_data}
\begin{tabular}{lcc}
\hline\hline
\multicolumn{1}{l}{S$_{\rm MW}$} & \multicolumn{1}{c}{JD - 2444000}&\multicolumn{1}{c}{W} \\ \hline
 0.223  &  -523.2  & 15956.8  \\
 0.207   &  -522.2  & 16877.2  \\
 0.234   &  -457.2  & 7264.7  \\
 0.206   &  -219.2  & 17620.2 \\
 0.218   &  -175.2  & 17201.9 \\
 0.213   &  -172.2  & 17475.1 \\
 0.219   &  -107.2  & 34054.6 \\
 0.205   &   172.8  & 8716.2  \\
 0.227   &   190.8  & 17363.2  \\
 0.221   &  200.8  & 10277.2  \\
 0.214   &   224.8   & 8639.2  \\
 0.205   &   260.8   & 8593.6  \\
 0.217   &   278.8   & 8700.7  \\
 0.194   & 482.9939 & 3409.7 \\
 ... &  ... & ... \\
\hline
\end{tabular}
\tablecomments{This table is published in its entirety in the machine-readable format. A portion is shown here for guidance regarding its form and content.}
\end{table*}

\bibliographystyle{aa}
\bibliography{manuscript}

\end{document}